\documentclass[a4paper,fleqn,usenatbib]{mnras}
\usepackage[T1]{fontenc}
\usepackage{ae,aecompl}

\usepackage{graphicx}	
\usepackage{amssymb}	

\usepackage[usenames]{color}

\newcommand {\bc}{\begin{center}}
\newcommand {\ec}{\end{center}}
\newcommand {\be}{\begin{equation}}
\newcommand {\ee}{\end{equation}}
\newcommand {\beq}{\begin{eqnarray}}
\newcommand {\eeq}{\end{eqnarray}}

\newcommand {\ergs}{{\rm erg\ \rm s^{-1}}}

\newcommand {\comment}[1]{}

\renewcommand{\d}{{\rm d}}

\title[Radiation beaming in bright XRPs]
{On the radiation beaming of bright X-ray pulsars and constraints on neutron star mass-radius relation}
\author[A. A.~Mushtukov et al.] 
{Alexander~A.~Mushtukov,$^{1,2,4}$\thanks{E-mail: al.mushtukov@gmail.com (AAM)}  
Patrick~A.~Verhagen,$^{1}$
Sergey S. Tsygankov,$^{3,4}$
\newauthor
Michiel van der Klis,$^{1}$
Alexander A. Lutovinov$^{4}$	 
and Tatiana I.	Larchenkova$^{4}$ \\
$^1$ Anton Pannekoek Institute, University of Amsterdam, Science Park 904, 1098 XH Amsterdam, The Netherlands \\
$^2$ Pulkovo Observatory, Russian Academy of Sciences, Saint Petersburg 196140, Russia \\
$^3$ Tuorla observatory, Department of Physics and Astronomy, University of Turku,
  V\"ais\"al\"antie 20, FI-21500 Piikki\"o, Finland \\  
$^4$ Space Research Institute of the Russian Academy of Sciences, Profsoyuznaya Str. 84/32, Moscow
  117997, Russia\\
} 
\date{Accepted 2017 September ??. Received 2017 September ??; in original form 2017 July ??}

\pubyear{2017}

\begin{document}
\label{firstpage}
\pagerange{\pageref{firstpage}--\pageref{lastpage}}
\maketitle

\begin{abstract}
The luminosity of accreting magnetised neutron stars can largely exceed the Eddington value due to appearance of accretion columns. The height of the columns can be comparable to the neutron star radius. The columns produce the X-rays detected by the observer directly and illuminate the stellar surface, which reprocesses the X-rays and causes additional component of the observed flux. The geometry of the column and the illuminated part of the surface determines the radiation beaming. Curved space-time affects the angular flux distribution. We construct a simple model of the beam patterns formed by direct and reflected flux from the column. We take into account the possibility of appearance of accretion columns, whose height is comparable to the neutron star radius. We argue that depending on the compactness of the star the flux from the column can be either strongly amplified due to gravitational lensing, or significantly reduced due to column eclipse by the star. The eclipses of high accretion columns result in specific features in pulse profiles. Their detection can put constraints on the neutron star radius. We speculate that column eclipses are observed in X-ray pulsar V~0332+53, leading us to the conclusion of large neutron star radius in this system ($\sim 15\,{\rm km}$ if $M\sim 1.4M_\odot$). We point out that the beam pattern can be strongly affected by scattering in the accretion channel at high luminosity, which has to be taken into account in the models reproducing the pulse profiles.
\end{abstract}


\section{Introduction}

X-ray pulsars (XRPs) are X-ray sources powered by accretion onto highly-magnetized neutron stars (NS) in binary systems. The magnetic field strength at the NS surface in these objects is typically $\gtrsim 10^{12}\,{\rm G}$ \citep{2015A&ARv..23....2W}. Such a strong magnetic field affects geometry of accretion flow and the fundamental properties of interaction between radiation and matter \citep{2006RPPh...69.2631H,2014PhyU...57..735P,2016PhRvD..93j5003M}, which shapes the basic properties of XRPs. The magnetic field was shown to be dominated by a dipole component in a few XRPs \citep{2016A&A...593A..16T}, however in some cases there is an evidence of strong non-dipole components of the field \citep{2017A&A...605A..39T}. The magnetic field channels the accretion flow to small areas ($\sim 10^{10}\,{\rm cm^2}$) on the NS surface, where the material loses its kinetic energy, which is emitted mostly in X-ray energy band. Misalignment of rotational and magnetic axis results in the phenomenon of X-ray pulsations. 

The geometry of the emitting region is defined by the magnetic field structure and the mass accretion rate. The accretion process results in hot spots (or low accretion mounds of height $\lesssim 100\,{\rm m}$, see e.g. \citealt{2013MNRAS.430.1976M,2013MNRAS.435..718M}) on the NS surface at relatively low mass accretion rates ($\lesssim 10^{17}\,{\rm g\,s^{-1}}$). However, at high mass accretion rates the radiation pressure becomes strong enough to stop accretion flow above NS surface \citep{2015MNRAS.447.1847M}, which leads to appearance of accretion columns (this critical luminosity has been recently detected by \citealt{2017MNRAS.466.2143D}). The material is stopped at the top of accretion column at a radiation dominated shock and then settles down to the NS surface. The matter in accretion column is confined by the strong magnetic field. Moreover, the strong magnetic field can significantly reduce  the scattering cross section \citep{1979PhRvD..19.2868H,2016PhRvD..93j5003M} which determines the radiation pressure. Under these conditions the system can produce a luminosity well above the Eddington limit, which is $L_{\rm Edd}\simeq 2\times 10^{38}\,\ergs$ for a typical NS. This concept can explain ultraluminous XPRs \citep{2015MNRAS.454.2539M}, whose luminosity is detected to be as high as $10^{40}-10^{41}\,\ergs$ \citep{2014Natur.514..202B,2017Sci...355..817I,2017MNRAS.466L..48I}. 

The geometry of the emitting regions affects the observational manifestations of X-ray pulsars: their spectral \citep{2015MNRAS.454.2714M,2013ApJ...777..115P,2015MNRAS.452.1601P} and timing properties \citep{2015MNRAS.448.2175L}. Particularly, the geometry defines the radiation beam pattern and, therefore, the observed pulse profiles \citep{1973A&A....25..233G}. The observational manifestation can also be affected by the accretion flow at the magnetospheric surface, which tends to be optically thick at extremely high mass accretion rates $\gtrsim 10^{20}\,{\rm g\,s^{-1}}$ \citep{2017MNRAS.467.1202M}.

NSs are extremely compact objects, whose radius exceeds the gravitational radius $r_{\rm s}=2GM/c^2$ by a factor of few only, and they strongly affect the space-time geometry in their vicinity. Thus, the effects of general relativity (GR) have to be taken into account in order to reconstruct the observational properties of XRPs. Typical XRPs are slowly rotating objects with spin period above $1$ second (e.g. \citealt{2015A&ARv..23....2W}). In this case the curved space-time is well described by the Schwarzschild metric. 

Previously, pulse profiles affected by GR effects were constructed for the case of hot spots on the NS surface \citep{1983ApJ...274..846P,2006MNRAS.373..836P,2010A&A...520A..76A}, which is valid for low mass accretion rates. The pulse profiles at high but still sub-critical mass accretion rates can be affected by the non-trivial initial beam pattern due to the scattering in the accretion channel \citep{1975A&A....42..311B} and asymmetry of its base \citep{1995ApJ...450..763K}.	

Formation of a beam pattern for the case of accretion columns at high mass accretion rates was also considered \citep{1978A&A....70..133M,1988ApJ...325..207R,2001ApJ...563..289K}, but the columns were assumed to be small ($H<1\,{\rm km}$). However, it was recently shown that the columns above magnetized NSs can be as high as the NS radius ($\sim 10\,{\rm km}$, see e.g. \citealt{2013ApJ...777..115P}). It also was ignored in theoretical models that the radiation intercepted by NS surface is reprocessed (reflected) and contributes to the total observed flux (though, the reprocessed component was mentioned in the interpretation of the decomposed X-ray signal from bright XRPs V~0332+53 and 4U 0515+63, see e.g. \citealt{2012A&A...540A..35S}). It is important to note that the total X-ray flux can even be dominated by reflected component within certain range of super-critical accretion luminosities \citep{2013ApJ...777..115P,2015MNRAS.448.2175L}. 

In this paper we construct a simplified model which describes beaming of the X-ray flux from super-critical XRPs, where the initial photon energy flux originates from the accretion column. The total flux is composed of direct flux from the column and the reflected signal from the NS surface. GR effects results in strong light bending and in lensing of X-ray flux from the accretion column by the NS. The lensing might result in a very high photon energy flux in directions opposite to the accretion columns. It is interesting that accretion columns of sufficient height cannot be completely eclipsed by the NS because of GR light bending. The  possibility of the eclipses is defined by accretion column height, which depends on the mass accretion rate and $B$-field strength, and compactness of a NS. We argue that the eclipsing manifests itself by a sharp dip in the X-ray pulse profile and that its detection at a certain accretion luminosity provides an upper limit on NS compactness (or a lower limit on NS radius for a given NS mass).

We also point out that in the case of high mass accretion rate the accretion flow at the magnetospheric surface tends to be optically thick \citep{1976SvAL....2..111S} and can influence the beam pattern, especially along the magnetic field axis. At extremely high mass accretion rates typical for the recently discovered pulsating ULXs \citep{2014Natur.514..202B,2017Sci...355..817I,2017MNRAS.466L..48I} the accretion flow at the magnetospheric surface forms an optically thick envelope, which shapes the observed pulse profiles \citep{2017MNRAS.467.1202M}. We discuss the influence of the magnetospheric accretion flow and show that it can dramatically change the observational manifestation of super-critical XRPs.

\section{Basic ideas}
 
\begin{figure}
\centering 
\includegraphics[width=7.1cm]{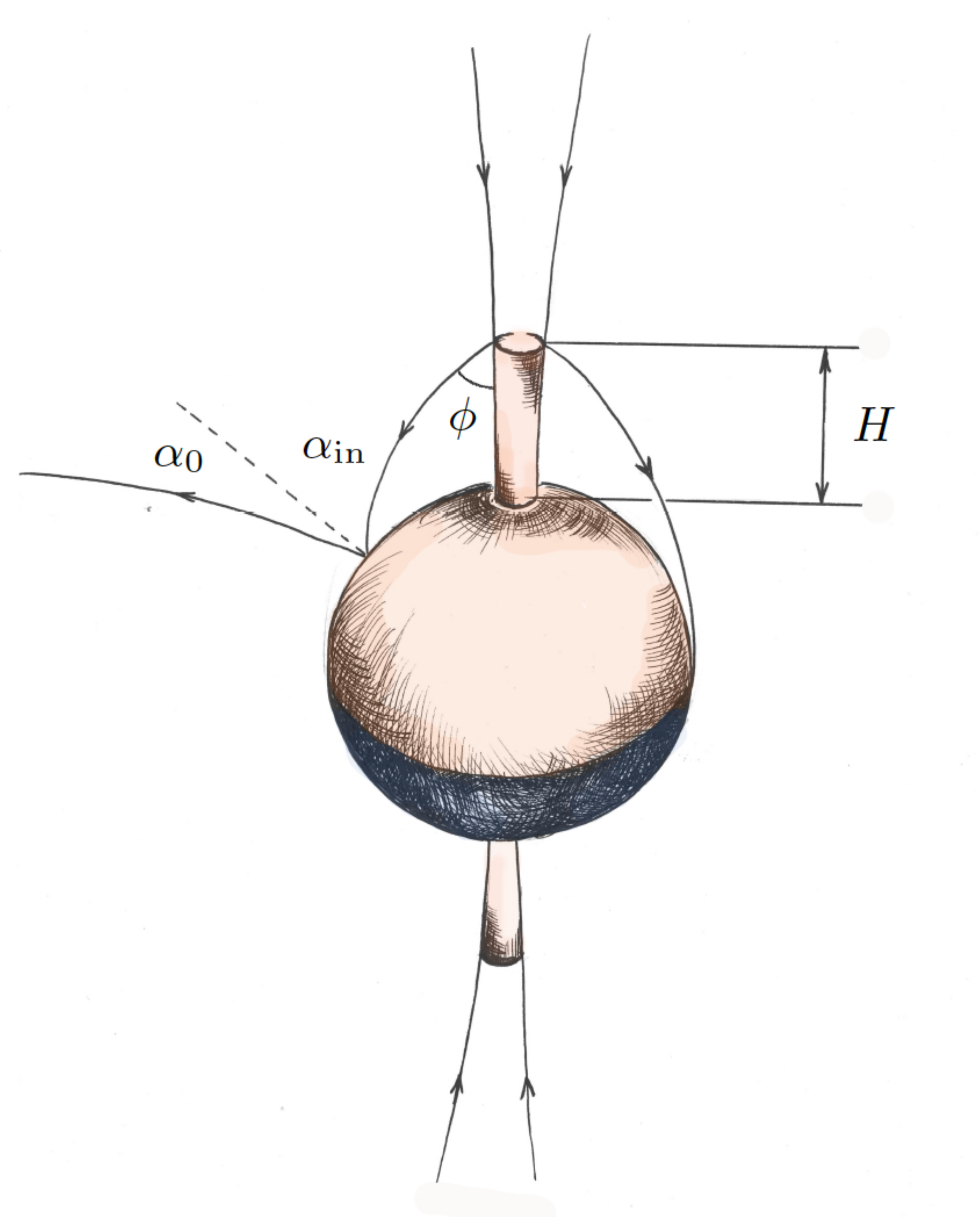} 	
\caption{The scheme of super-critical accreting NS illuminated by accretion column of height $H$. The observer detects the X-ray photon energy flux directly from the columns and the flux reflected by the NS surface. The composition of the two components defines the observed variability of a source.}
\label{pic:scheme}
\end{figure}

We consider a spherically symmetric magnetized NS of mass $M$ and radius $R$ with a geometrically thin accretion column above its surface (see Fig.\,\ref{pic:scheme}).  The height of the accretion column $H$ depends on the mass accretion rate $\dot{M}$ and can be comparable to the NS radius \citep{BS1976,2015MNRAS.454.2539M}. The radiation of the accretion column is beamed towards the NS surface due to photon scattering by fast electrons at the edge of accretion channel \citep{1976SvA....20..436K,1988SvAL...14..390L}. Thus, the NS intercepts a significant fraction of the total luminosity of the accretion column \citep{2013ApJ...777..115P}. The fraction of the intercepted radiation is even higher if one takes into account effects of GR (light bending). 
The intercepted radiation is reprocessed (reflected) by the NS surface and contributes to the total flux of the object. 
As a result, the photon flux from a source is composed of direct flux from the accretion column and flux reflected by the atmosphere of NS.

Accreting material forms an envelope around the NS located at the surface of magnetosphere. The mechanism of opacity in the envelope depends on its local temperature and mass density. The envelope can be optically thick in a regions close to NS magnetic poles. At sufficiently high accretion luminosity (typical to recently discovered pulsating ULXs, $\sim 10^{40}\,\ergs$) the envelope is optically thick everywhere and fully reprocesses the initial radiation from the central object \citep{2017MNRAS.467.1202M}. 

The magnetic field of a NS is considered to be dominated by dipole component, which is likely a case for classical XRPs (see \citealt{2016A&A...593A..16T}).

\subsection{Accretion luminosity and accretion column height}

The geometry of accretion channel and accretion column are defined by the magnetic field structure and radius of NS magnetosphere
\be 
R_{\rm m}=2.5\times 10^{8}\Lambda B_{12}^{4/7}L_{37}^{-2/7}M_{1.4}^{1/7}R_6^{10/7} \,\,\,{\rm cm},
\ee
where $\Lambda<1$ is a constant with $\Lambda=0.5$ being a commonly used value for the case of accretion from a disc \citep{2014EPJWC..6401001L}, $B_{12}=B/10^{12}\,{\rm G}$ is the magnetic field strength at the NS surface, $L_{37}=L/10^{37}\,\ergs$ is the accretion luminosity, $M_{1.4}=M/(1.4M_\odot)$ is the NS mass, ans $R_6=R/10^6\,{\rm cm}$ is the NS radius. The radius of the base of accretion channel can be roughly estimated as
\beq
r_{\rm b}&\simeq& R\left(\frac{R}{R_{\rm m}}\right)^{1/2} \nonumber \\
&\simeq& 6.3\times 10^4\,\Lambda^{-1/2}B_{12}^{-2/7}L_{37}^{1/7}M_{1.4}^{-1/14}R_6^{11/14}\,\,{\rm cm}.
\eeq
The relation between accretion column height and luminosity is defined by the accretion channel base geometry and opacity across $B$-field lines $\kappa_\perp$, which depends on magnetic field strength. There is an approximate relation between column height and luminosity \citep{2015MNRAS.454.2539M}:
\be\label{eq:H2L}
L\approx 38\left(\frac{l_0/d_0}{50}\right)\left(\frac{\kappa_{\rm T}}{\kappa_\perp}\right)f\left(\frac{H}{R}\right)L_{\rm Edd},
\ee
where
\be 
f\left(\frac{H}{R}\right)\equiv \log\left(1+\frac{H}{R}\right)-\frac{H}{R+H}\nonumber,
\ee
$l_0\sim 2\pi r_{\rm b}\sim 5\times 10^5\,{\rm cm}$ and $d_0\sim 10^3 - 10^4\,{\rm cm}$ are geometrical length and thickness of the accretion channel at the NS surface and $\kappa_{\rm T}\approx 0.34\,{\rm cm^2\,g^{-1}}$ is the opacity due to non-magnetic Compton scattering.
The height where the shock wave arises varies within the accretion channel \citep{1988SvAL...14..390L}. As a result, the geometrical thickness of a sinking region $x$ is given by
\be
\frac{x}{d/2}=\left(1-\frac{h}{H}\frac{R+H}{R+h}\right)^{1/2} 
\ee
and the optical thickness of the free-fall region due to the Thomson scattering can be estimated as
\be
\tau_{\rm ff}=\frac{1.7\times 10^{5}}{l_0}\frac{L_{37}}{\beta_{\rm ff}}\left(1-\frac{x}{d/2}\right),
\ee
where $\beta_{\rm ff}=v_{\rm ff}/c<1$ is the dimensionless free-fall velocity. We see that the optical thickness depends on the mass accretion rate (or accretion luminosity) and height above NS surface. For given mass accretion rate, $\tau_{\rm ff}\propto 1/l_0$. Large optical thickness of the free-falling region results in strong beaming of the X-ray radiation from accretion column walls \citep{1976SvA....20..436K,1988SvAL...14..390L,2013ApJ...777..115P}.

\subsection{Problem in a flat space-time}

\subsubsection{Point source above NS surface}

\begin{figure}
\centering 
\includegraphics[width=7.cm]{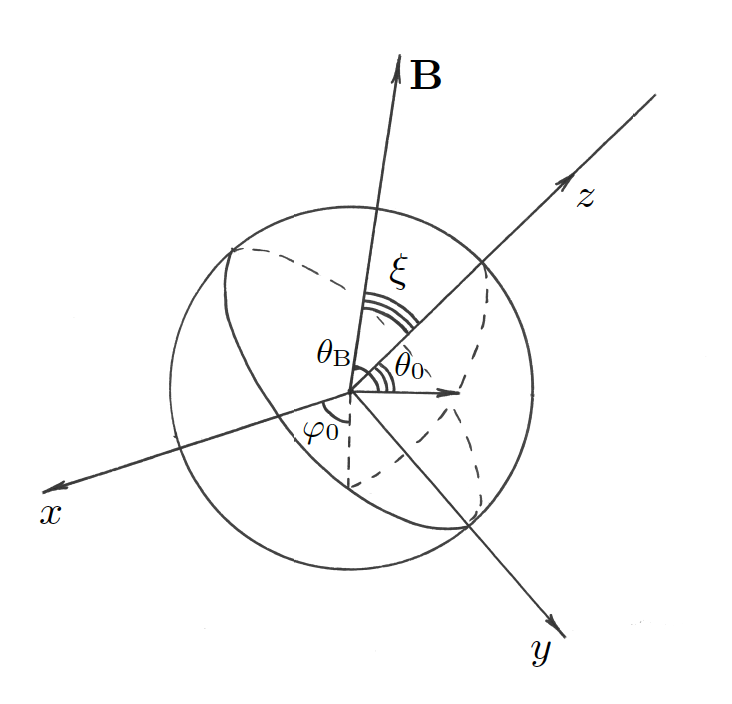} 	
\caption{The coordinate angles in the observer's ($\theta_0$ and $\varphi_0$) and $B$-field reference frames ($\theta_{\rm B}$). $\xi$ is the angle between the line of sight, which is aligned with $z$-axis, and the $B$-field axis.}
\label{pic:geom_scheme01}
\end{figure}

Lets consider the simplified problem of an isotropic point source of total luminosity $L_{\rm ps}$ located at a height $h$ above the NS surface. The space-time is considered to be flat in this section. Then the incoming photon energy flux $F_{\rm in}$ at the NS surface is given by
\be\label{eq:surFlux1}
F_{\rm in}(\theta_{\rm B})=\left(\frac{L_{\rm ps}}{4\pi D(\theta_{\rm B},h)^2}\right)\cos\alpha_{\rm in}(\theta_{\rm B},h), 
\ee
where
\beq
D(\theta_{\rm B},h)=\sqrt{h^2+2R(R+h)(1-\cos\theta_{\rm B})}\nonumber
\eeq
is a distance from the point source to the point at the NS surface defined by latitude $\theta_{\rm B}$ in $B$-field reference frame (see Fig.\,\ref{pic:geom_scheme01}) and
\beq
\cos\alpha_{\rm in}=\frac{h\cos\theta_{\rm B}-R(1-\cos\theta_{\rm B})}{D} \nonumber
\eeq
defines the angle between local  normal to the NS surface and photon momentum (see Fig.\,\ref{pic:scheme}).
The incoming flux is reprocessed by the NS surface and emitted back in to space. In a stationary model the local incoming photon energy flux is equal to the local emitted flux:
\be
F_{\rm in}(\theta_{\rm B})=2\pi\int\limits_{0}^{\pi/2}\d \alpha_0 I_{\rm out}(\theta_{\rm B},\alpha_0)\cos \alpha_0\sin \alpha_0, 
\ee 
where $I_{\rm out}(\alpha_0)$ is the intensity of the reflected radiation, and $\alpha_0$ is the angle between the local normal to the NS surface and the photon momentum (see Fig.\,\ref{pic:scheme}). If the intensity of the reflected radiation does not depend on $\alpha_0$, we get
$$
F_{\rm in}(\theta_{\rm B})=\pi I_{\rm out}(\theta_{\rm B}).
$$
Integrating over the visible part of the NS surface, we get the photon energy flux, which is detected by a distant observer from the surface of the illuminated NS:
\be\label{eq:F_obssur_flat_01}
F_{\rm obs, sur} \propto \int\limits_{0}^{2\pi}\d\varphi_0\int\limits_{0}^{\pi/2}\d\theta_0 \sin \theta_0\cos \theta_0 I(\theta_0,\varphi_0,\alpha_0),
\ee
where $\theta_0$ and $\varphi_0$ are coordinate angles at the NS surface in the observer reference frame (see Fig.\,\ref{pic:geom_scheme01}). In case of flat space-time $\alpha_0=\theta_0$ and 
\beq 
I(\theta_0,\varphi_0,\alpha_0)=I_{\rm out}(\theta_{\rm B},\alpha_0).
\eeq
The latitude $\theta_{\rm B}$ at the NS surface in the $B$-field reference frame can be obtained from coordinates in the observer's reference frame by the relation
\beq
\cos\theta_{\rm B}=\sin\xi\sin \varphi_0\sin\theta_0+\cos\xi\cos\theta_0,
\eeq 
where $\xi$ is the angle between the magnetic field axis and the line of sight (see Fig.\,\ref{pic:geom_scheme01}).

In case of a non-isotropic but axisymmetric distribution of the photon energy flux from the point source, the equation (\ref{eq:surFlux1}) should be rewritten in a more general form:
\be
 F_{\rm in}(\theta_{\rm B})=\left(\frac{L_{ps}f(\phi)}{4\pi D(\theta_{\rm B},h)^2}\right)\cos\alpha_{\rm in}(\theta_{\rm B},h), 
\ee
where function $f(\phi)$ describes the angulal distribution of radiation, and $\phi$ is the angle between the direction from point source to NS center and photon momentum. The distribution function $f(\phi)$ satisfies the normalization:
$2\pi\int_{0}^{\pi}\d\phi f(\phi) \sin\phi =1$.

\subsubsection{Radiation from the accretion column}

The accretion column is an extended source of X-ray radiation defined by its height $H$, the size of its base, and the distribution of the emitted flux over height and directions.

\begin{figure}
\centering 
\includegraphics[width=8.5cm]{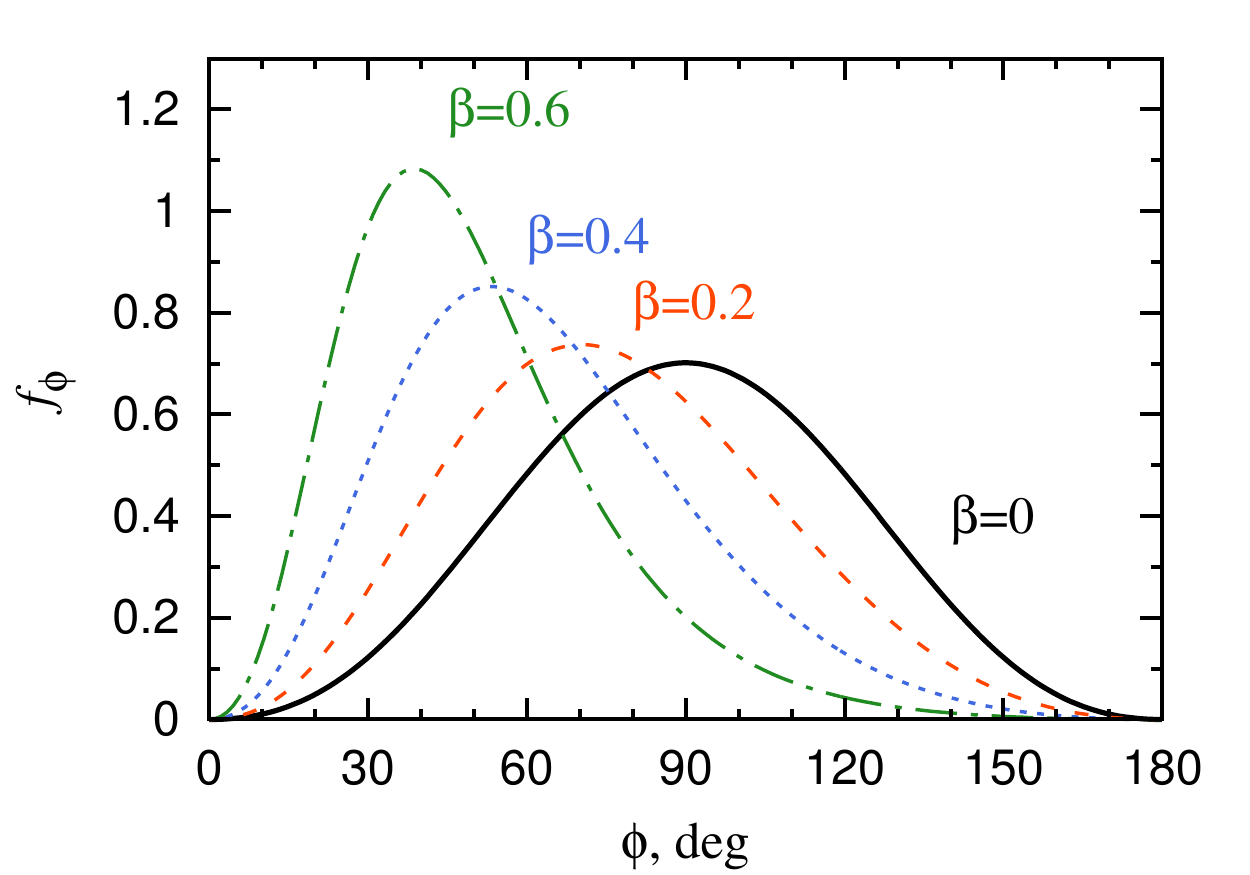} 	
\caption{Normalized distribution of radiation over the directions $f_\phi$ at the accretion column wall. Different curves are given for different dimensionless velocities $\beta=0$ (solid black), $0.2$ (dashed red), $0.4$ (dotted blue) and $0.6$ (dashed-dotted green). One can see that the radiation is strongly beamed towards NS surface in case of high $\beta$.}
\label{pic:LubarskiiDiag}
\end{figure}

Let us use a function which describes the angular distribution of the emitted power: $\frac{\d L(\phi)}{\d \phi}$, which is normalized by 
$$
\int\limits_{0}^{\pi}\d \phi\frac{\d L(\phi)}{\d \phi}=L_{\rm tot}.
$$
In case of a point source and flat space-time the flux distribution over the surface is
\be
F_{\rm in}(\theta_{\rm B})=\left(\frac{\d L(\phi)/\d \phi}{2\pi\sin\phi D(\theta_{\rm B},h)^2}\right)\cos\alpha_{\rm in}(\theta_{\rm B},h), 
\ee
where $\phi=\alpha_{\rm in} - \theta_{\rm B}$. Further it would be convenient to use another normalization for the angular distribution of emitted power:
\be
f_\phi \equiv \frac{\d L(\phi)/\d \phi}{L_{\rm tot}},
\ee
then the normalization is $\int_{0}^{\pi}\d\phi f_\phi =1 \nonumber.$
The angular distribution of emitted power is defined by the velocity of the accretion flow at the edges of the accretion channel $\beta=v/c$ at given height:
\be
\left(\frac{\d L}{\d \phi}\right)= \frac{I_0 \sin^2\phi}{\gamma^5(1-\beta\cos\phi)^4}
\left(1+\frac{\pi}{2}\frac{\sin\phi}{\gamma(1-\beta\cos \phi)}\right),
\ee
where $I_0$ is a normalization constant and $\gamma\equiv(1-\beta^2)^{-1/2}$ \citep{2013ApJ...777..115P}. Because 
\be
\int\limits_{0}^{\pi}  \frac{I_0 \sin^2\phi \left(1+\frac{\pi}{2}\frac{\sin\phi}{\gamma(1-\beta\cos \phi)}\right)\,\d\phi}{\gamma^5(1-\beta\cos\phi)^4}=\frac{7\pi I_0}{6} \nonumber,
\ee
the normalized angular power distribution is given by (see Fig.\,\ref{pic:LubarskiiDiag})
\be\label{eq:RadBeam}
f_\phi= \frac{6 \sin^2\phi \left(1+\frac{\pi}{2}\frac{\sin\phi}{\gamma(1-\beta\cos \phi)}\right)}{7\pi \gamma^5(1-\beta\cos\phi)^4}.
\ee
The velocity at the edges of accretion channel is likely close to the local free fall velocity $\beta\simeq \beta_{\rm ff}=(r_{\rm s}/(R+h))^{1/2}$ (see e.g. \citealt{1988SvAL...14..390L}).

The distribution of emitted power over the height in accretion column is given by
\be
g(h)\equiv \frac{\d L_{\rm acc}(h)}{\d h},\quad\quad \int\limits_{0}^{H}\d h\,g(h)=L_{\rm acc}, 
\ee
where $L_{\rm acc}$ is the total luminosity of the accretion column. 
The exact distribution of emitted power over the height is provided by models of accretion column. In this paper we use the  distribution described by
\be
g(h)\propto \frac{1}{R+H}\frac{H-h}{R+h}, 
\ee
which was derived for accretion columns in a diffusion approximation \citep{2015MNRAS.454.2539M}.
Because the accretion column can be considered as a set of point sources of luminosity $g(h)\d h$ and the distribution of incoming photon energy flux over the surface can be calculated as 
\be\label{eq:F_in}
 F_{\rm in}(\theta_{\rm B})\propto \int\limits_{h_{\rm min}}^{H}\d h\,\left(\frac{f_\phi (\phi(\theta_{\rm B},h)) g(h)}
 {2\pi D(\theta_{\rm B},h)^2 \,\sin\phi(\theta_{\rm B},h)}\right)\cos\alpha_{\rm in}(\theta_{\rm B},h),
\ee
where $h_{\rm min}$ is the minimal accretion column height contributing to the flux at the latitude $\theta_{\rm B}$ ($h_{\rm min}=R(1-\cos\theta_{\rm B})/\cos\theta_{\rm B}$ in the case of flat space-time).
As soon as we know the flux distribution over the NS surface, we can calculate the reflected flux which is detected by a distant observer (see eq.\,(\ref{eq:F_obssur_flat_01})).

Using the luminosity distribution over the column height and the local angular power distribution we get the flux which is detected by the observer directly from the accretion column:
\be
F_{\rm obs, col}\propto \frac{1}{2\pi\sin\phi'}\int\limits_{h_{\rm min}}^{H}\d h f_\phi(\phi')\, {g(h)},
\ee
where the integration performed over the visible part of a column, $\phi'=\xi$ or $\phi'=\pi-\xi$ depending on the orientation of given accretion column. Note that the visible parts are generally different for two accretion columns (for example, one accretion column can be totally visible, while the other is partly or totally eclipsed by NS).

\subsection{Effects of general relativity}

The photon propagation in the vicinity of a NS is affected by the gravitational field. In order to take the effects of GR into account we consider photon propagation in the Schwarzschild metric, which corresponds to a non-rotating central object. This approximation is valid for XRPs, whose spin periods $P$ are typically about a few seconds or longer. The photon trajectories in case of the Schwarzschild metric are described by the differential equation \citep{1973grav.book.....M}: 
\be\label{eq:PhTrajectory}
\frac{\d^2 u}{\d\varphi^2}+u=3u^2, 
\ee
where $u\equiv 0.5\,r_{\rm s}/r$, $r$ is the radial coordinate of the photon, and $\varphi$ is the angle between the photon momentum and the radius vector directed from the NS center to the current position of the photon. Equation (\ref{eq:PhTrajectory}) can be solved numerically for given initial location $r_0$ (which gives $u_0$) and momentum of a photon (which gives $(\d u/d\varphi)_0=-u_0/\tan\alpha$, where $\alpha$ is the angle between radius-vector and photon momentum).

\subsubsection{The direct photon energy flux from the accretion column and illumination of the NS surface}

The photon energy flux detected by a distant observer directly from the accretion column is affected by the initial radiation beaming at the edge of the column walls, gravitational light bending and orientation of the NS in the observer's reference frame because the photons from the accretion column can be intercepted by the NS and the accretion column can be partly or totally eclipsed by the NS. 

The observed photon flux from the visible part of the accretion column is given by 
\be
F^{\rm (GR)}_{\rm obs, col}\propto \frac{1}{2\pi}\int\limits_{h_{\rm min}}^{H}\d h
\frac{{g(h)}f_\phi(\phi')}{\sin\phi'},
\ee
where $h_{\rm min}$ is the minimum height of a point above the NS surface, which is not eclipsed by the NS in a given orientation in the observer's reference frame (if $h_{\rm min}>H$, then $F^{\rm (GR)}_{\rm obs, col}=0$),
the angle $\phi'$ depends on height $h$ and orientation of a NS described by angle $\xi$ (see Fig.\,\ref{pic:geom_scheme01}):
$$
\cos\phi' \simeq \pm \frac{r_{\rm s}}{R+h}+\left(1-\frac{r_{\rm s}}{R+h}\right)\cos\xi,
$$
where the sign on the right hand side of the equation depends on the accretion column considered: the one in the front of the NS or the one on the back side of the NS in the observer's reference frame.

The photon energy flux $F_{\rm in}$ which is intercepted and reprocessed by the NS surface at given latitude $\theta_B$ is given by equation (\ref{eq:F_in}), where angles $\alpha_{\rm in}$ and $\phi$ should be recalculated to take into account light bending. We recalculated the angles numerically, but one can use the approximate relations: 
\be
\sin\alpha_{\rm in}=\sin\phi\,\frac{R+h}{R}\sqrt{\frac{1-r_{\rm s}/R}{1-r_{\rm s}/(R+h)}},\nonumber
\ee
\be 
\cos\phi = \frac{r_{\rm s}}{R+h}+\left(1-\frac{r_{\rm s}}{R+h}\right)\cos\psi,\nonumber
\ee
\be 
\cos\alpha_{\rm in} = \frac{r_{\rm s}}{R}+\left(1-\frac{r_{\rm s}}{R}\right)\cos(\psi+\theta_{\rm B}),\nonumber
\ee
where the angle $\psi$ defines the direction of photon momentum at the infinity.
These approximate expressions for the angles $\alpha_{\rm in}$, $\phi$ and $\phi'$ are based on approximations proposed by \cite{2002ApJ...566L..85B}, which are not very accurate for NSs of extreme compactness. In our case we use numerical solutions of the differential equation (\ref{eq:PhTrajectory}) in order to get accurate values of the angles.

\begin{figure}
\centering 
\includegraphics[width=8.5cm]{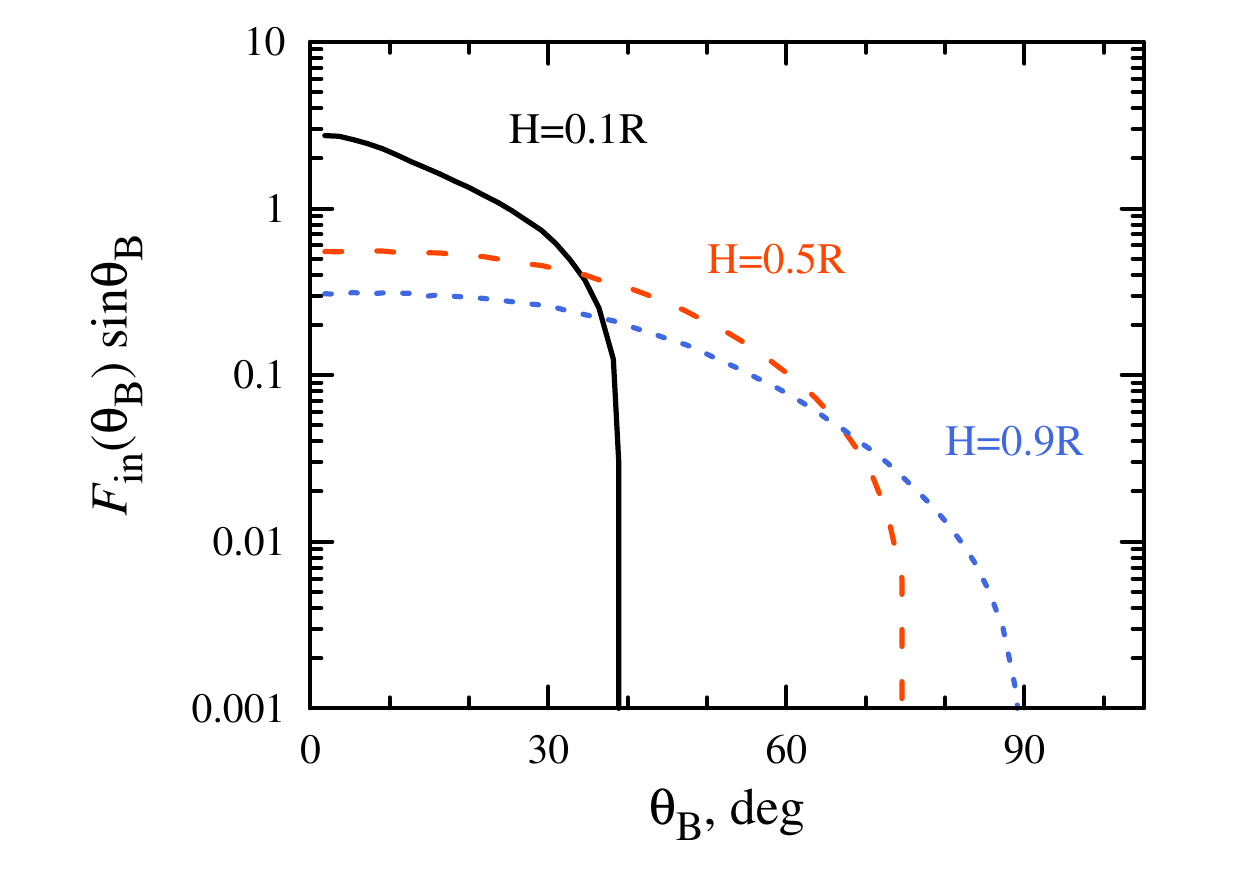} 	
\caption{The distributions of photon energy flux over the NS surface illuminated by an accretion column of height $H=0.1R$ (black solid line), $0.5R$ (red dashed line) and $0.9R$ (blue dotted line). The distribution of the initial photon energy flux over  height in the accretion column is taken to be uniform. Parameters: $M=1.4\,M_\odot$, $R=10\,{\rm km}$.}
\label{pic:FluxOnTheSurface}
\end{figure}

\subsubsection{Photons from the NS surface}

Because of curved photon trajectories, the observer can detect photons originating from more than half the NS surface. The maximum observed latitude in the observer's reference frame $\theta_{0,{\rm max}}\geq \pi/2$ (see Fig.\,\ref{pic:geom_scheme01}) can be roughly estimated as
\be
\theta_{0,{\rm max}}\simeq \arccos\left( -\frac{r_{\rm s}}{R}(1-r_{\rm s}/R)^{-1} \right),
\ee
which gives $\theta_{0,{\rm max}}\sim 136^\circ$ and corresponds to $\sim 86$ per cent of the total NS surface for a NS of mass $M=1.4\,M_\odot$ and radius $R=10\,{\rm km}$.

The photons which can be detected by the observer from latitude $\theta_0$ (see Fig.\,\ref{pic:geom_scheme01}) are emitted from the NS surface in a direction given by angle $\alpha_0$ (see Fig.\,\ref{pic:scheme}). One can get the approximate relation between $\alpha_0$ and $\theta_0$ \citep{2002ApJ...566L..85B}:
\be
\cos\alpha_0 \simeq \frac{r_{\rm s}}{R}+\left(1-\frac{r_{\rm s}}{R}\right)\cos\theta_0,
\ee
where $\theta_0<\theta_{\rm 0,max}$
Then the observed flux from the NS surface can be obtained by integration over the visible part of the NS:
\be 
F^{\rm (GR)}_{\rm obs, sur}\propto
\int\limits_{0}^{2\pi}\d\varphi_0\int\limits_{0}^{\theta_{0,{\rm max}}}\d\theta_0 \sin \theta_0\cos \alpha_0 I_{\rm out}(\theta_0,\varphi_0,\alpha_0),
\ee
where $\varphi_0$ is the longitude on the NS surface in the observer's reference frame and the intensity of the radiation emitted in a given direction $I_{\rm out}$ is defined by the local incoming photon flux from the accretion column.
In our numerical calculations the intensity of the reprocessed radiation at the NS surface is taken to be constant at every $\alpha_0<\pi/2$. This does not affect the qualitative results, but it would be necessary to include the actual angular distribution of intensity in order to get accurate predictions for the shape of the pulse profiles. The law of X-ray reflection can strongly dependent on photon energy, polarization state, local strength and direction of the $B$-field. This problem is beyond the scope of this paper.

\section{Influence of the accretion curtain}

The accreting matter, which is moving along magnetic field lines from accretion disc to the NS surface, can affect the observed X-ray flux due to absorption and scattering processes in it \citep{1976SvAL....2..111S}. The optical thickness of the envelope is determined by the mass accretion rate and the mechanism of opacity. 

Kramers' opacity is defined by the mass density $\rho$ of the accretion flow and its temperature $T_{\rm g}$. For the case of pure hydrogen Kramers' opacity is given by
\be
\kappa_{\rm a}=0.0136\,\rho T^{-3.5}_{\rm g,keV}\,{\rm cm^2\,g^{-1}},
\ee
but it can vary significantly with the chemical composition of the accreting material. 
The opacity due to Compton scattering is given by $\kappa_{\rm e}=0.34\,{\rm cm^{2}\,g^{-1}}$ in the case of non-magnetic scattering. 

The temperature of the accretion flow at the magnetosphere is determined by the energy release due to interaction between accretion disc and NS magnetic field. It can be roughly estimated from the known mass accretion rate and the NS spin period $P$ \citep{2017MNRAS.467.1202M}: 
\be
T\simeq 0.3\,\frac{\gamma_{\rm a}-1}{1+X}\Lambda^{-1}L_{37}^{2/7}B^{-4/7}_{12}m^{6/7}R_6^{-10/7}
\left[1-\frac{\Omega}{\Omega_{\rm K}}\right]^2\,\,{\rm keV},\nonumber
\ee
where $\gamma_{\rm a}$ is the adiabatic index, $X$ is the hydrogen mass fraction, $\Omega=2\pi/P$ is the angular velocity of the magnetosphere and $\Omega_{\rm K}$ is the Keplerian angular velocity at the magnetosphere.
At high mass accretion rate ($\dot{M}\gtrsim 10^{17}\,{\rm g\,s^{-1}}$) the temperature is high and opacity is dominated by Compton scattering. Kramers' opacity can be important at relatively low mass accretion rates ($\dot{M}\lesssim 10^{17}\,{\rm g\,s^{-1}}$), as has been detected e.g. in XRP RX J0440.9+4431, where the dip-like structure in pulse profile was observed at energies below 8 keV \citep{2012MNRAS.421.2407T}.  

A strong magnetic field results in resonant scattering of photons, whose energy is close to the local cyclotron energy $E^*_{\rm cyc}\approx 11.6\,B^*_{12}\,{\rm keV}\approx 11.6\,B_{12}(R/(R+h))^3\,{\rm keV}$. The cross-section of resonant Compton scattering exceeds the non-magnetized scattering cross-section by a few orders of magnitude \citep{2016PhRvD..93j5003M}. Because the magnetic field decreases with distance from the NS surface ($B\propto (R+h)^{-3}$ in case of a dipole magnetic field), the photons with energies below the cyclotron energy at the NS surface $E_{\rm cyc}$ can be resonantly scattered at the appropriate height \citep{1986ESASP.251..375Z}. Resonant scattering leads to a complex beam pattern and, thus, we can speculate that the relatively complex pulse profiles at low energies and simple smooth pulse profiles at high energies of the majority of XRPs are explained by the influence of resonant scattering at $E\lesssim E_{\rm cyc}$. It is interesting that XRPs with relatively low surface magnetic field strength (e.g. GRO J1744-28, where the surface magnetic field strength was reported to be $\sim 10^{11}\,{\rm G}$, see e.g. \citealt{2015MNRAS.452.2490D}) show simple pulse profiles over the entire X-ray energy band.

In order to estimate the influence of scattering in the accretion envelope we use the opacity of non-magnetized Compton scattering. The optical thickness due to the scattering can be estimated as \citep{2017MNRAS.467.1202M}
\be\label{eq:tau01}
\tau_{\rm e}(\lambda)
\approx\frac{1.4\,L^{6/7}_{37}B^{2/7}_{12}}{\beta(\lambda)}\left(\frac{\cos\lambda_0}{\cos\lambda}\right)^3,
\ee
where $\beta$ is local dimensionless velocity of the accretion flow, $\lambda$ is the coordinate angle measured from the equator of the magnetic dipole and $\lambda_0$ is the coordinate angle, which corresponds to the edge of a polar cap at the NS surface.

\begin{figure}
\centering 
\includegraphics[width=8.5cm]{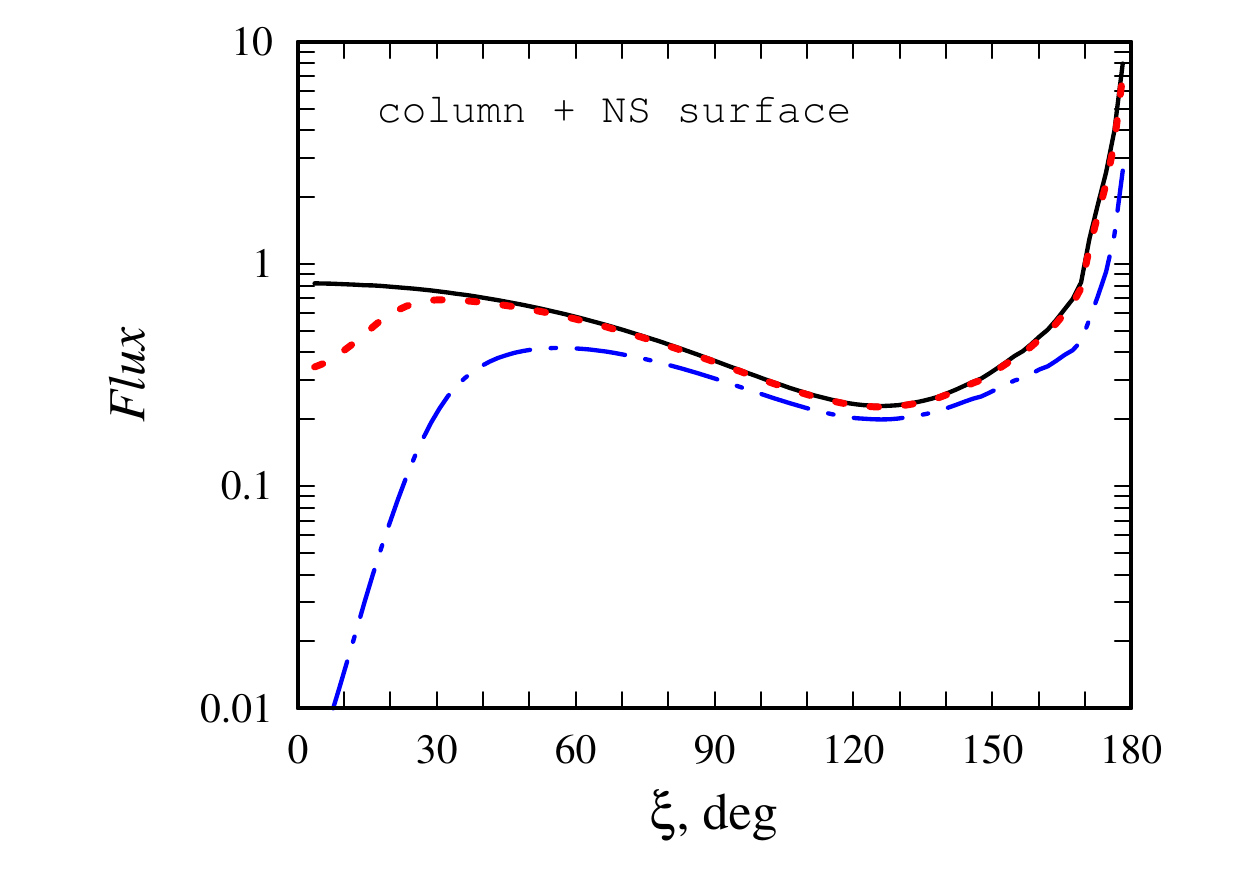} 	
\caption{Distribution of non-scattered flux over angle $\xi$. 
{Different curves are given for different mass accretion luminosities (and, therefore, different optical thickness of the curtain): $10^{37}\,\ergs$ (black solid line), $10^{38}\,\ergs$ (red dotted line) and $10^{39}\,\ergs$ (blue dashed-dotted line).}
The influence of scattering is stronger along the magnetic field axis, where the optical thickness of the accretion flow is higher.
{One can see that photon scattering in the accretion curtain may result in dips in the pulse profile if the mass accretion rate in high enough.}
Parameters: $H=0.5R$, $M=1.4\,M_\odot$, $R=10\,{\rm km}$.}
\label{pic:absorption}
\end{figure}

At the high mass accretion rates relevant to super-critical accretion ($\gtrsim 10^{37}\ergs$, see e.g.  \citealt{2015MNRAS.447.1847M}) the optical thickness of the envelope can be of order unity or even higher. 
Numerically solving equation (\ref{eq:PhTrajectory}) we get the point where photons originating from the accretion column or NS surface cross the accretion flow at the magnetosphere surface. Taking into account the local optical thickness (\ref{eq:tau01}), we can estimate the fraction of non-scattered intensity: $I/I_0\approx e^{-\tau_{\rm e}}$. Then we can roughly estimate the influence of the envelope on the beam pattern formation (see Fig.\,\ref{pic:absorption}). The photons scattered by accretion envelope are redistributed over all directions and affect the observed pulse profile. However, their influence is beyond the scope of this paper.

\begin{figure}
\centering 
\includegraphics[width=8.5cm]{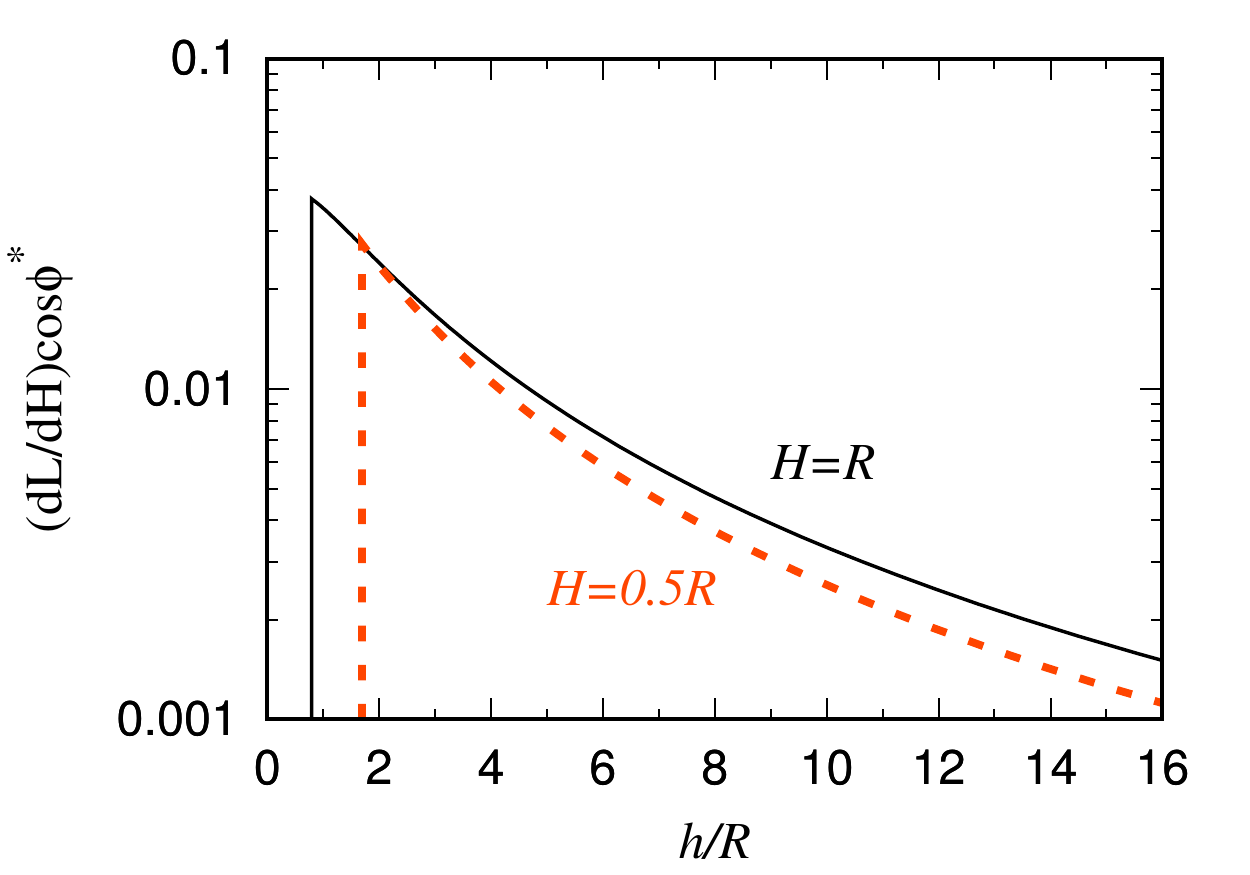} 	
\caption{{The distribution of radiation power of a point source (we multiply the power distribution with $\cos\phi^*$ in order to get the part of the power, which affects the dynamics along $B$-field lines), which crosses the magnetic axis on the back side of a NS relative to the column, over the height $h$ above the surface. Different curves correspond to different height of the source above the surface: $H=R$ (black solid line) and $R=0.5R$ (red dashed line). Parameters: $R=2.4r_{\rm s}$. The initial beaming of the radiation is described by equation (\ref{eq:RadBeam}).}}
\label{pic:CrossX_2}
\end{figure}

\begin{figure}
\centering 
\includegraphics[width=8.5cm]{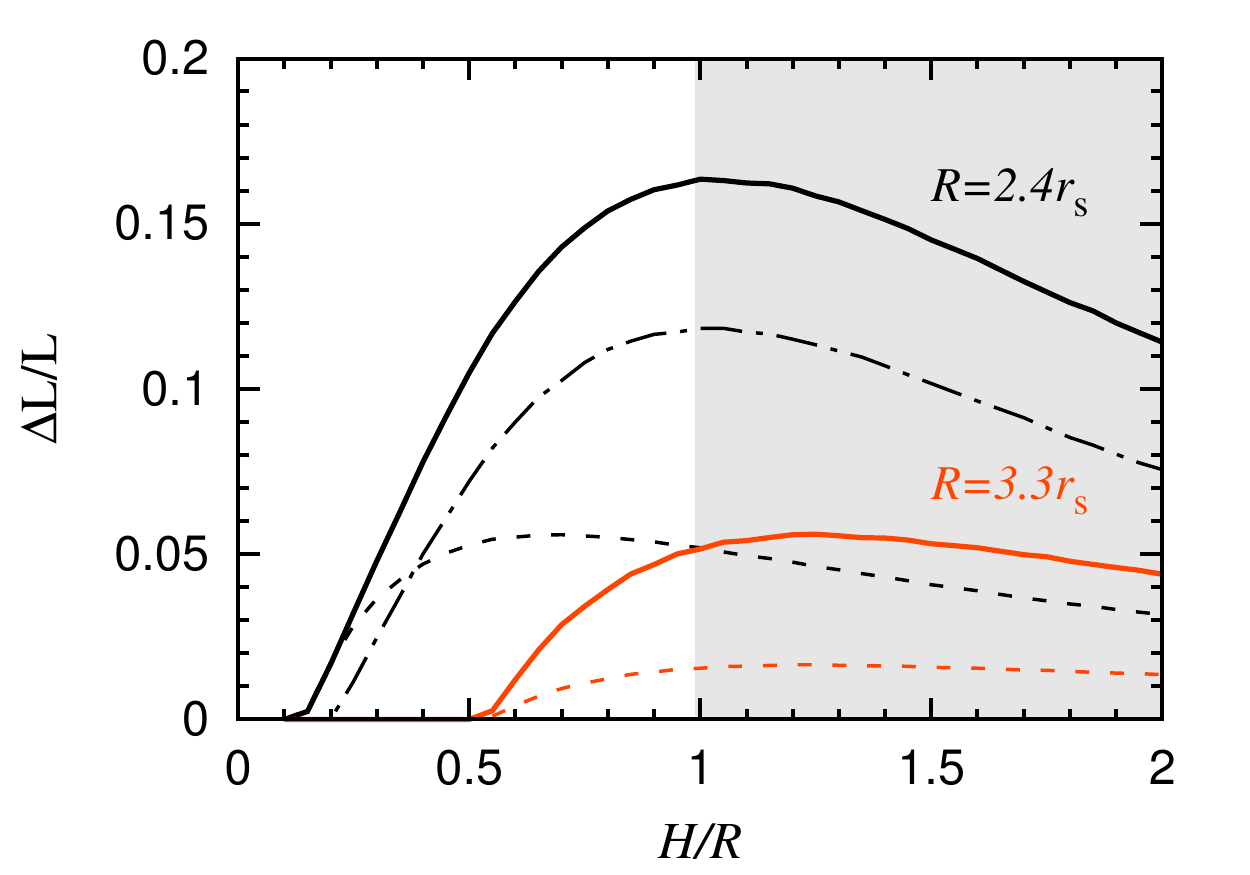} 	
\caption{The fraction of point source luminosity which crosses the magnetic axis on the opposite side of the NS, as a function of relative height $(H/R)$ of a source above the surface. Different curves are given for various NS radii (black lines correspond to $R=2.4\,r_{\rm s}$, while red lines correspond to $R=3.3\,r_{\rm s}$) and initial beaming of radiation (solid lines correspond to the beaming caused by photon scattering by fast electrons given by equation (\ref{eq:RadBeam}), while dashed lines correspond to isotropic intensity at the walls of the accretion column). Black dashed-dotted line corresponds to the fraction of point source luminosity intercepted by the accretion channel within $10$ NS radii. Grey region on the plot corresponds to extremely big heights which we consider unlikely.}
\label{pic:CrossX}
\end{figure}

The radiation intercepted by the accretion channel can influence the accretion flow itself. At the heights typical for accretion columns ($H<R$), effects of irradiation by the opposite column are not significant unless the accretion column is as high as the NS radius and the NS is extremely compact (see Fig.\,\ref{pic:CrossX_2}, where the results are given for a NS of small radius $R=2.4r_{\rm s}$). Otherwise, the irradiation will affect the dynamics of accretion flow well above the shock region in the accretion column. 
One can estimate the fraction of the point source luminosity intercepted by the accretion channel, which can be as high as $5-15$ percent depending on the location of the source, the initial beaming of the radiation and the compactness of the NS (see Fig.\,\ref{pic:CrossX}). In case of an accretion column the influence of the intercepted radiation depends on the luminosity distribution over height in the accretion column as well.

\section{On the construction of pulse profiles}

\begin{figure}
\centering 
\includegraphics[width=7cm]{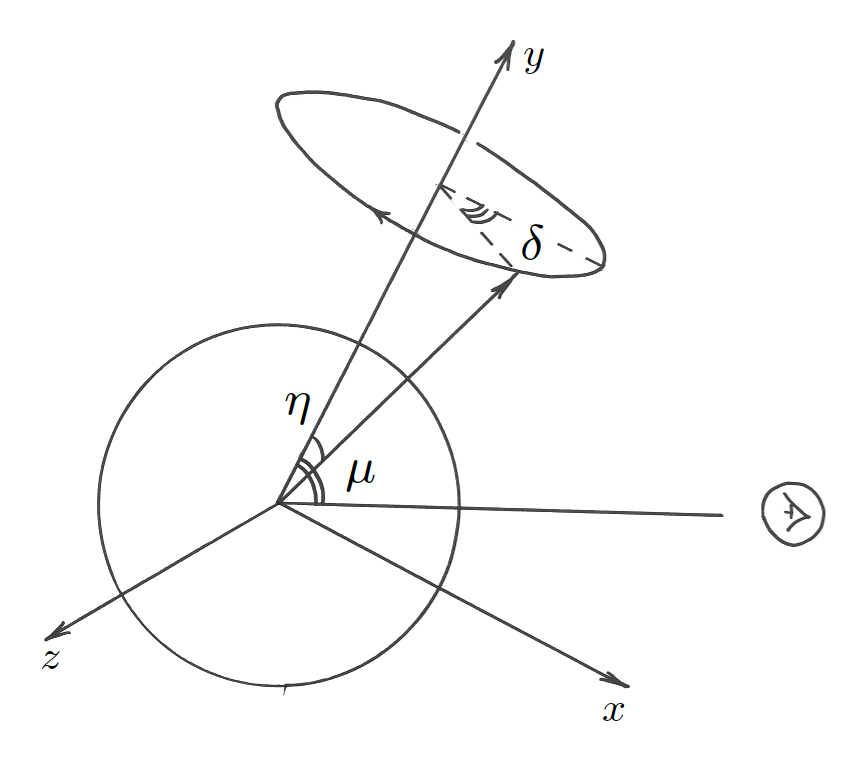} 	
\caption{The orientation of a NS in the observer's reference frame is defined by three angles: the angle between line of sight and rotational axis $\mu$, the angle between rotational axis and $B$-field axis $\eta$  and phase angle $\delta$.}
\label{pic:geom_scheme02}
\end{figure}

The pulse profiles can be constructed out of the known X-ray beam pattern (angular distribution of the total X-ray flux in the reference frame of the NS) and geometry of the rotating NS in the observer's reference frame (see Fig.\,\ref{pic:geom_scheme02}) given by: $\mu$, the angle between line of sight and rotational axis, $\eta$, the angle between rotational axis and $B$-field axis and $\delta$, the phase angle, which varies within the interval $[0;2\pi]$ during the pulse period. The angle $\xi$ between the line of sight and the magnetic axis  (see Fig.\,\ref{pic:geom_scheme01}), which determines the total photon energy flux detected by distant observer, is totally defined by $\mu$, $\eta$ and phase angle $\delta$:
\be
\cos\xi =\sin\mu\sin\eta\cos\delta +\cos\mu\cos\eta. 
\ee
One can see that the angle $\xi$ is variable within the range $[|\mu-\eta|,\,\mu+\eta]$. Fixing the angles $\mu$ and $\eta$ and taking various phase angles $\delta$ we can construct the pulse profile for a calculated beam pattern of XRPs.

Using the known pulse profile one can get the pulsed fraction (PF):
\be
{\rm PF}=\frac{F_{\rm max}-F_{\rm min}}{F_{\rm max}+F_{\rm min}}\leq 1, 
\ee
where $F_{\rm min}$ and $F_{\rm max}$ are minimum and maximum X-ray flux detected within the pulse period. Because the exact beam pattern might depend on the exact X-ray energy range \citep{2006MNRAS.371...19T}, the PF can be different for different bands. 

It is obvious that the PF contains less information about the XRP than the pulse profile. However, it is sensitive to the appearance of eclipses and further amplification of the observed flux from the opposite column due to effects of gravitational light bending. In that sense, the variations of PF with luminosity of XRP can be used as a diagnostic tool.

\begin{figure}
\centering 
\includegraphics[width=8.5cm]{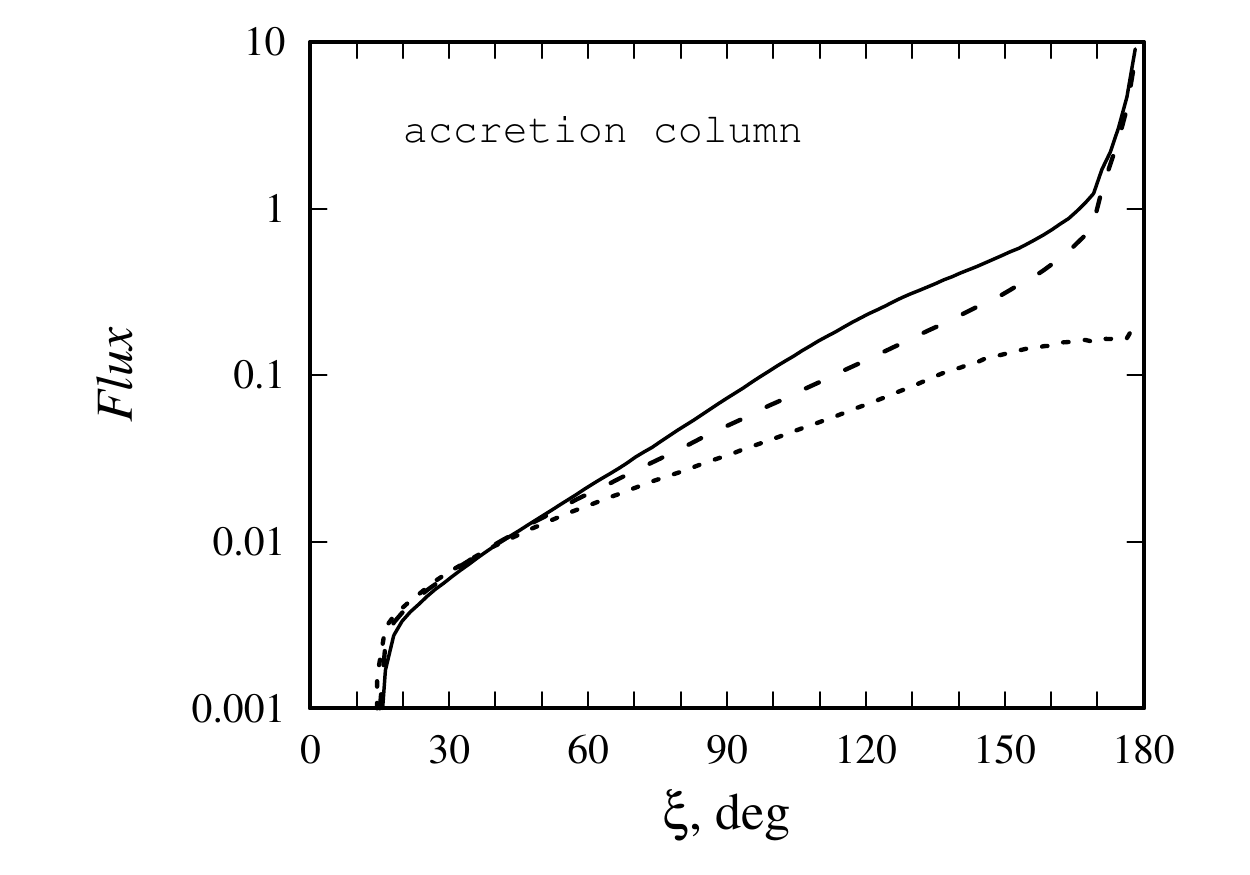} 	
\includegraphics[width=8.5cm]{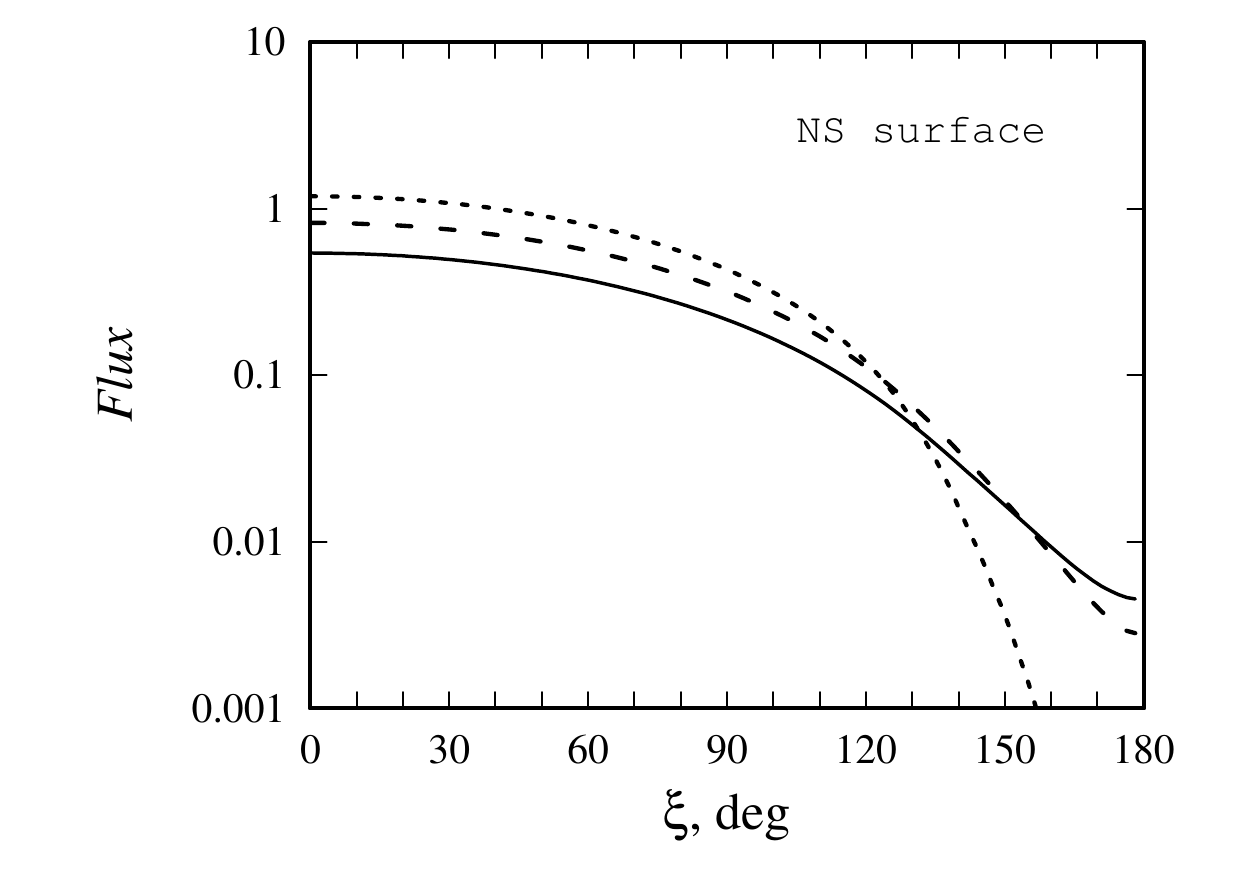} 	
\caption{Angular distribution of direct and reflected flux. Different curves are given for various relative heights of accretion column: $H=0.1,\,0.5,\,1R$ (dotted, dashed and solid lines, respectively). Parameters: $M=1.4\,M_\odot$, $R=10\,{\rm km}$.}
\label{pic:FlixColumnSurface}
\end{figure}

\begin{figure}
\centering 
\includegraphics[width=8.5cm]{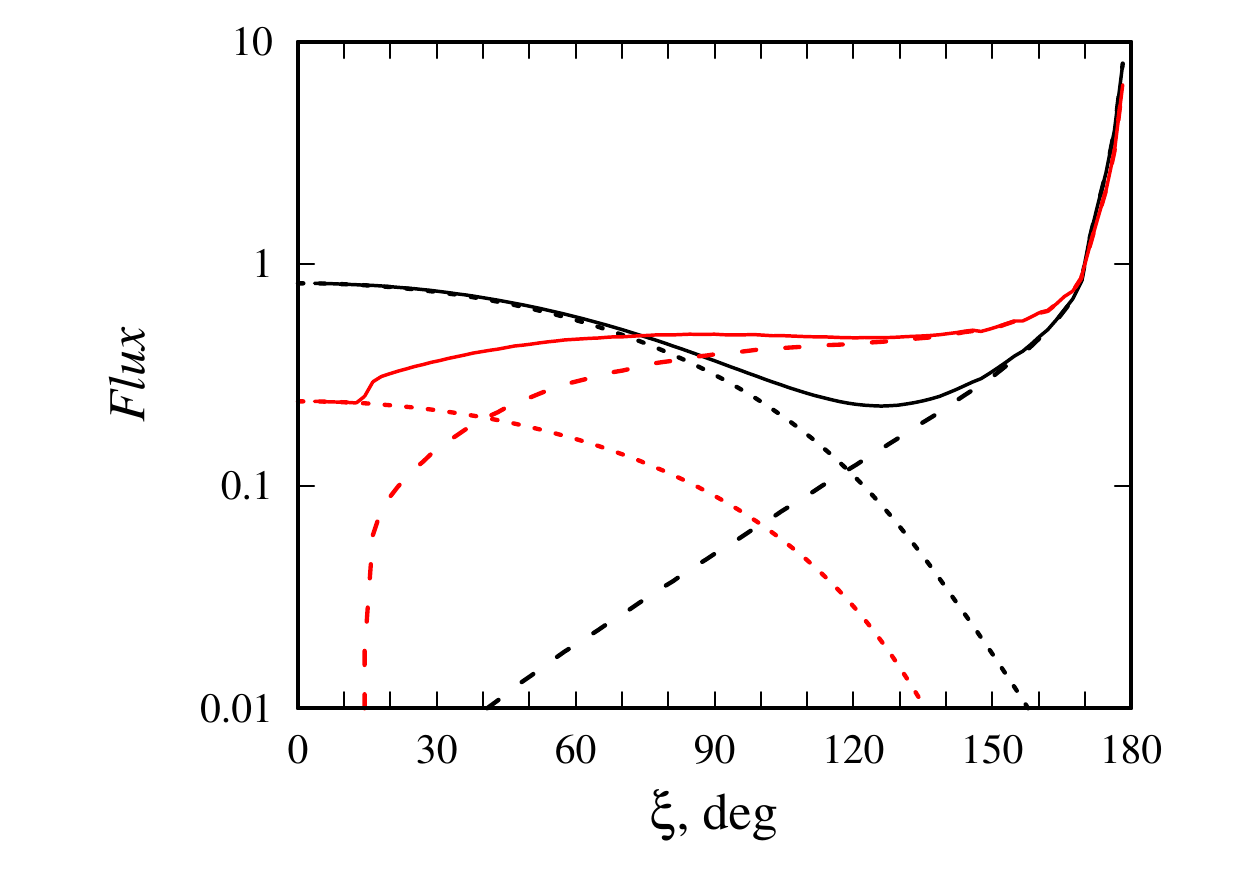} 	
\caption{The angular distribution of direct (dashed lines) and reflected (dotted lines) photon energy flux. The total photon energy flux is shown	 by solid lines. Black and red lines correspond to initially beamed (see eq. (\ref{eq:RadBeam})) and isotropic distributions of the photon energy flux at the walls of the accretion column. Parameters: $H=0.5R$, $r_{\rm s}=0.42$.}
\label{pic:diff_AC_beam}
\end{figure}

\begin{figure}
\centering 
\includegraphics[width=8.5cm]{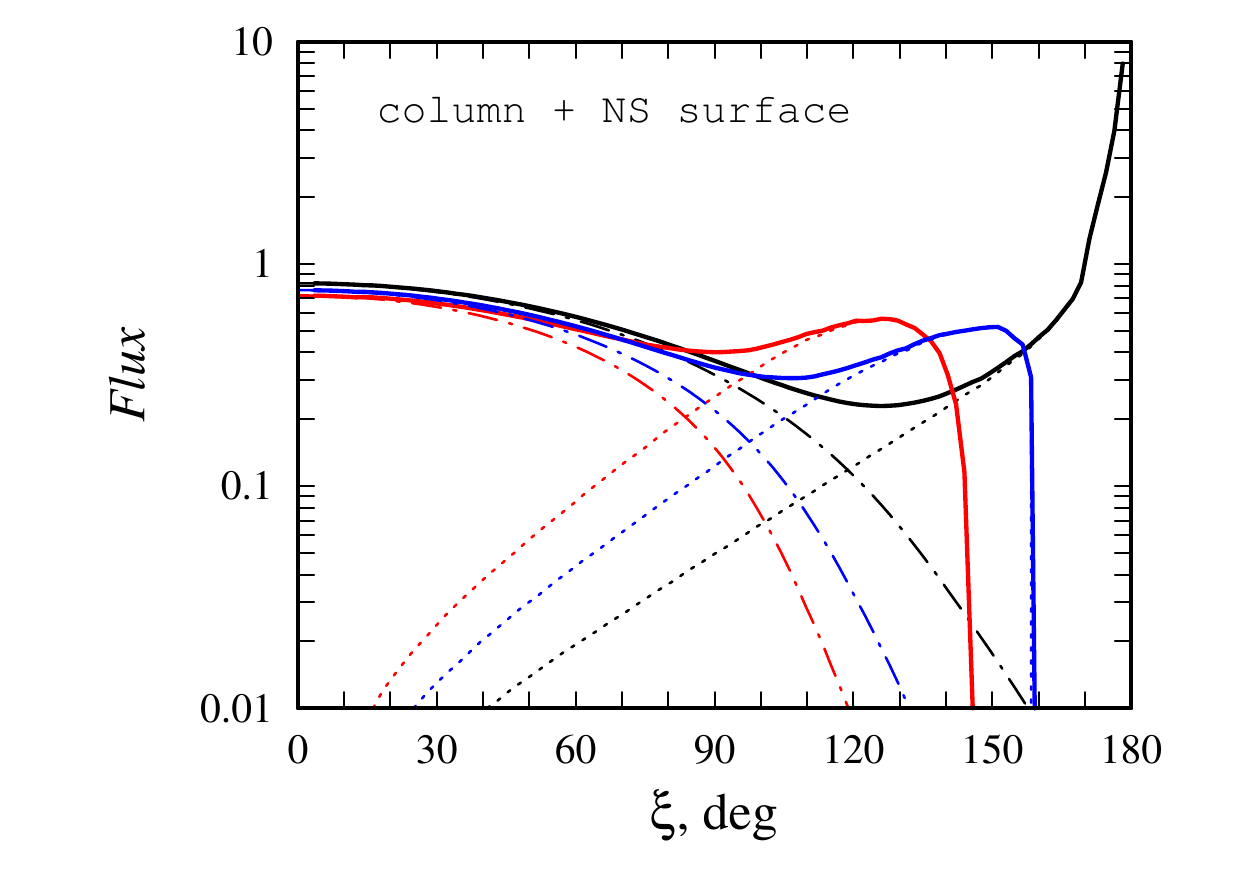} 	
\caption{Total angular flux distribution, given by solid lines for the cases of different compactness of a NS: $R_{\rm Sh}=0.42R$ (black), $R_{\rm Sh}=0.3R$ (blue), $R_{\rm Sh}=0.2R$ (red). Height of the accretion column is taken to be $H=0.5R$. Dotted and dashed-dotted lines give distribution of direct and reflected flux respectively.}
\label{pic:diff_compactness}
\end{figure}

\begin{figure}
\centering 
\includegraphics[width=8.5cm]{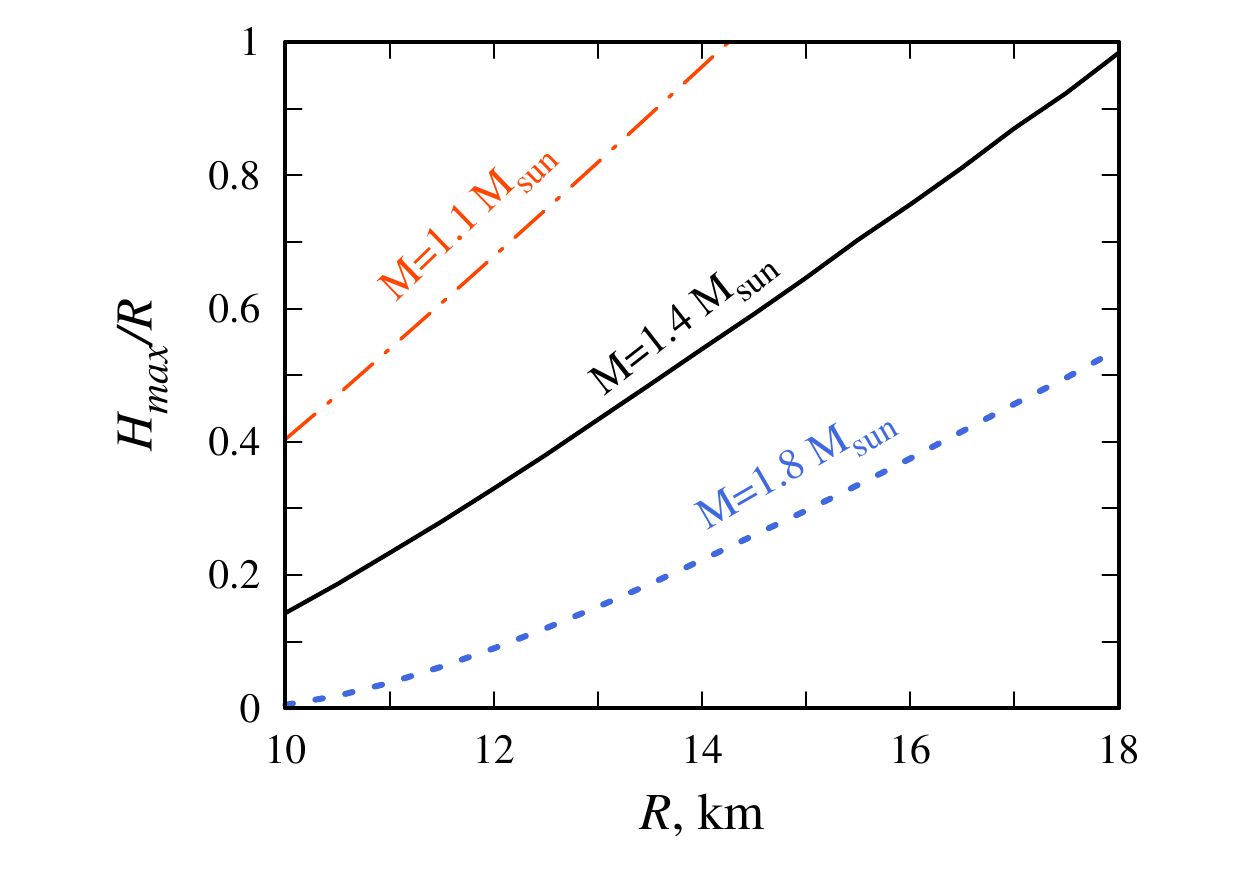} 	
\caption{The maximum relative height of the accretion column which can be entirely eclipsed by NS of a given radius. Different curves are given for different NS mass: $1.1\,M_\odot$ (red dashed-dotted), $1.4\,M_\odot$ (black solid), $1.8\,M_\odot$ (blue dotted).}
\label{pic:ColumnEcl}
\end{figure}

\begin{figure}
\centering 
\includegraphics[width=8.7cm]{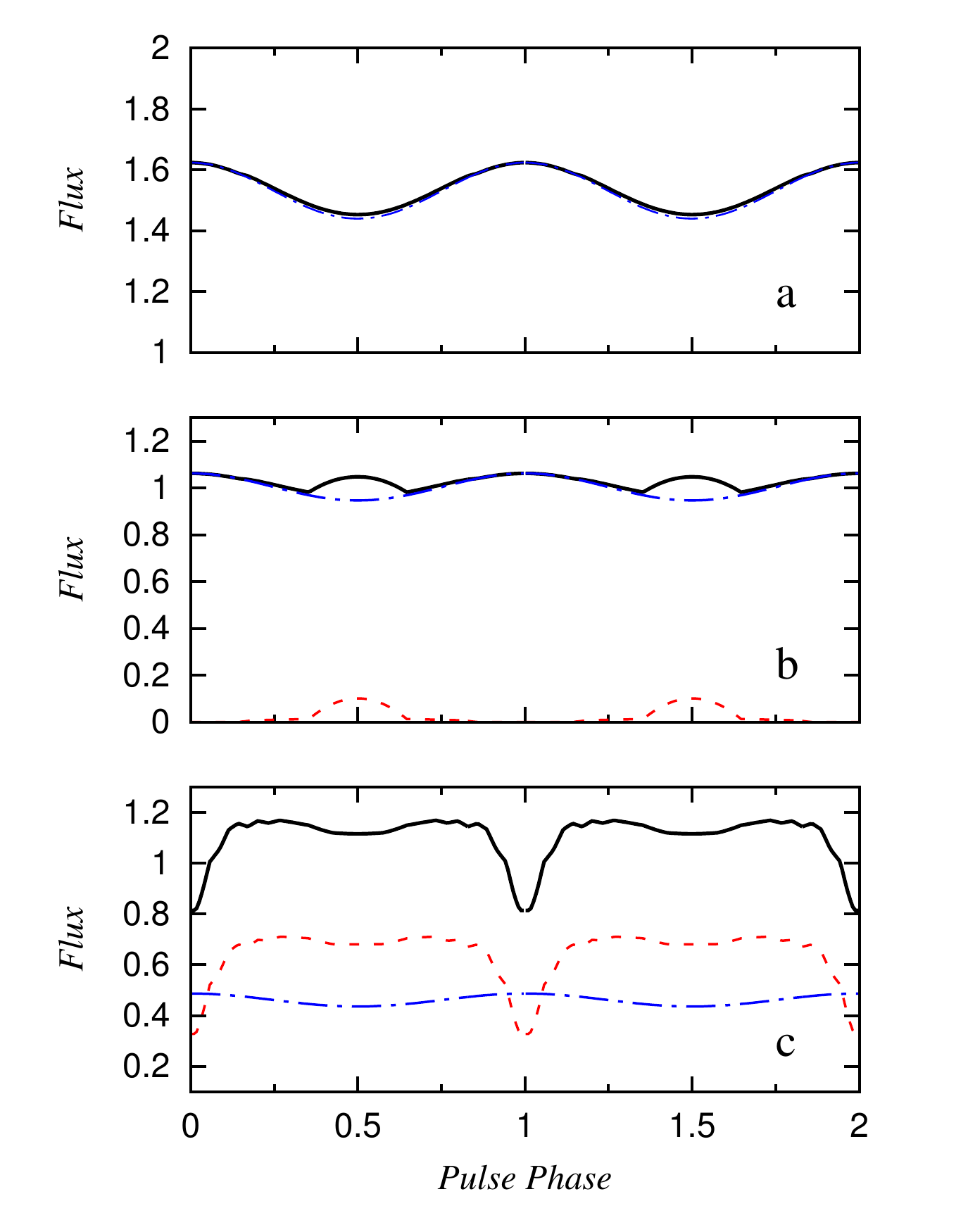} 	
\caption{Examples of modeled pulse profiles in (a) low-luminosity state, (b) intermediate luminosity state and (c) high luminosity state are shown by black solid lines. The contribution of flux from the accretion column and reflection from the NS surface are shown by red dashed and blue dashed-dotted lines correspondingly. At relatively low luminosity the flux is dominated by reflected signal, while at higher luminosity the contribution of the direct flux from the column is significant. The eclipses of the accretion column cause sharp dips in the observed pulse profile.}
\label{pic:sc_PP_3}
\end{figure}

\begin{figure}
\centering 
\includegraphics[width=8.5cm]{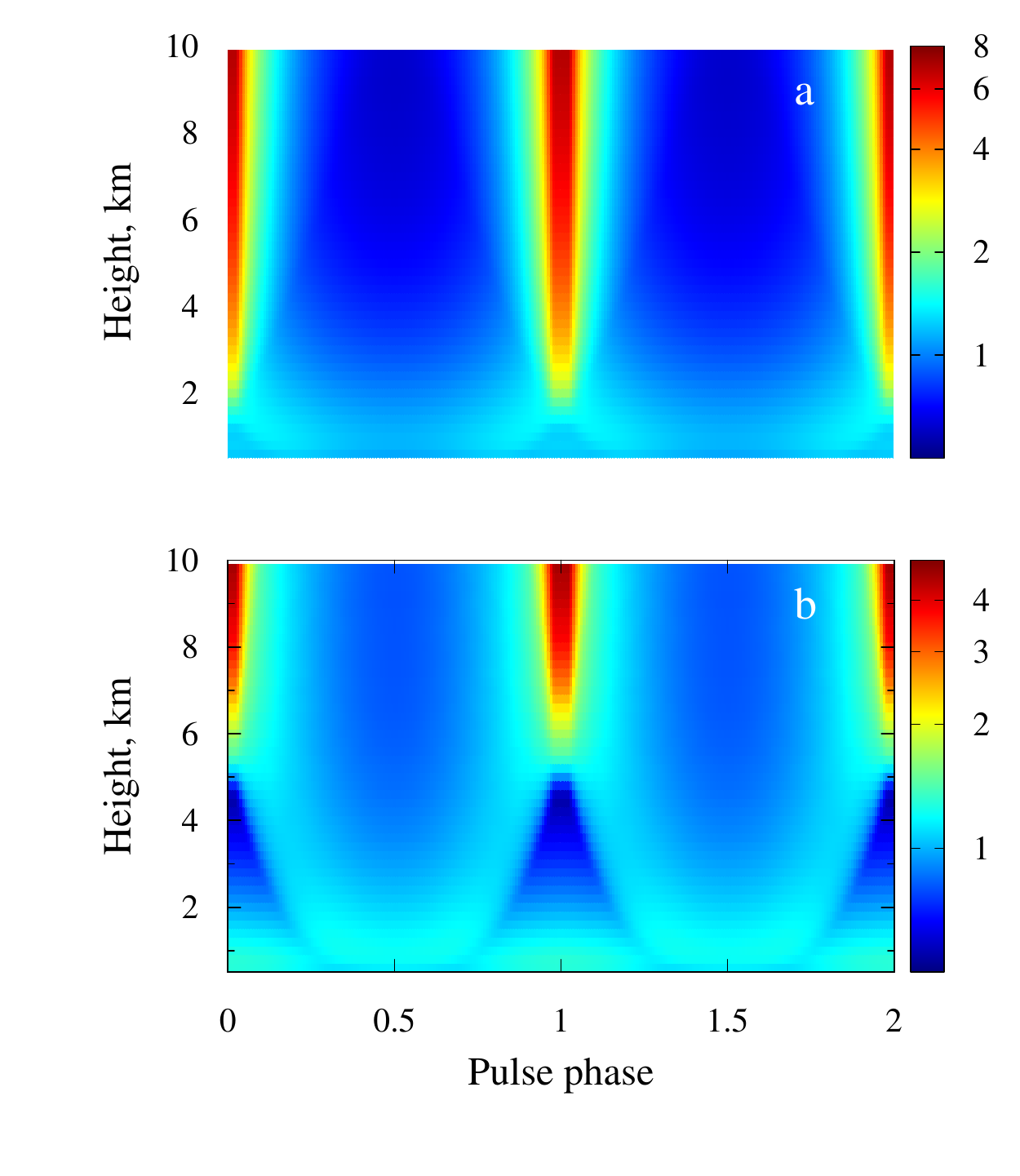} 
\caption{The variability of X-ray pulse profile with the height of accretion column. Two pictures are given for different parameters of a NS: (a) $M=1.4M_\odot,\,\,R=10\,{\rm km}$, (b) $M=M_\odot,\,\,R=10\,{\rm km}$. The scattering in the magnetospheric envelope is not taken into account.}
\label{pic:sc_pic23}
\end{figure}


\section{Discussion}

\subsection{Beam patterns of bright X-ray pulsars}

The beam pattern of a bright XRP is defined by the direct flux from the accretion column and the reflected flux from the NS surface (see Fig.\,\ref{pic:FlixColumnSurface}). Both of them depend on height of the accretion column, initial angular flux distribution at the accretion column walls (see Fig.\,\ref{pic:diff_AC_beam}) and NS compactness (see Fig.\,\ref{pic:diff_compactness}). 
In can be seen from these figures, that the direct flux from the accretion column is not distributed according to fan beam diagram, and moreover the flux distribution for a high accretion column is described better by a pencil beam diagram which is strongly peaked at the direction opposite to accretion column relative to the NS (see Fig.\,\ref{pic:diff_AC_beam}). 

It is interesting to note that observational evidence of a pencil beam from a high accretion column was already reported by \cite{2015MNRAS.448.2175L}, where the conclusions were based on the dynamics of a cyclotron line scattering feature detected in phase-resolved spectroscopy.
Depending on accretion column height and NS compactness, the direct flux from the accretion column is eclipsed, or strongly amplified due to gravitational lensing. High accretion columns, which are expected at high mass accretion rates \citep{BS1976,2015MNRAS.454.2539M}, can be eclipsed by NS of sufficiently large radii only (see Fig.\,\ref{pic:ColumnEcl}). Otherwise, the X-ray flux is strongly amplified in the direction opposite to the column relative to the NS. Therefore, evidence of eclipses of the accretion column by the NS can give a lower limit to the NS radius. 

However, the direct flux from the accretion column can be strongly reduced due to scattering by the accretion flow confined by the magnetic field in the accretion channel (see Fig.\,\ref{pic:absorption}).
{In the case of an axisymmetric accretion flow the influence of the scattering on the final pulse profile is stronger if the observer looks down at one of the NS magnetic poles. If the mass accretion rate is high enough ($\dot{M} > 10^{18}\,{\rm g\,s^{-1}}$), the scattering might result in dips in the observed pulse profile similar to those excepted from the eclipses.
In the case of a non-axisymmetric distribution of plasma on the NS magnetosphere, the dips caused by scattering can even appear  at lower mass accretion rates and not necessarily on top of local maximum of X-ray flux in pulse profile. A detailed analysis of the dips caused by scattering requires accurate calculations of the radiative transfer problem and knowledge of the exact shape of the accretion flow onto the magnetosphere, which is beyond the scope of this paper. }

The X-ray flux reflected from the NS surface forms a much more isotropic emission pattern than the direct flux from the accretion column. However, the reflected part of the beam pattern can be affected by scattering in the accretion channel as well.

Assuming a model of the beam pattern composed of two components: direct flux from accretion column and reflected flux from the NS surface, one can make qualitative predictions about the evolution of the PF over a range of the mass accretion rates. 

At mass accretion rates slightly above the critical value \citep{2015MNRAS.447.1847M} most of the X-ray flux is intercepted by the NS surface and beam pattern is defined by reflected component. The higher the mass accretion rate, the higher the accretion column, the bigger the illuminated part of NS surface, the smaller the variations of detected X-ray flux and the smaller the corresponding PF. 

At relatively high mass accretion rates the accretion column is high enough that only a small fraction of the X-ray flux is intercepted by the NS surface. The X-ray flux from the column is strongly beamed due to Compton scattering by relativistic electrons \citep{1988SvAL...14..390L} and due to gravitational lensing by NS. In this situation the higher the mass accretion rate, the higher the column, the larger the direct component of strongly beamed X-ray flux, the stronger the X-ray flux variability and the higher the PF.

Thus, one would expect a decrease of the PF with luminosity at  relatively low mass accretion rates and a further increase at sufficiently high mass accretion rates. The detailed dependence of the PF on accretion luminosity is defined by the exact orientation of the accreting NS in the observer's reference frame and the exact beaming of the initial X-ray flux, which might be different in different energy bands.

\subsubsection{Constrains on the NS radius in XRP V~0332+53}

If the accretion column is high enough, it can hardly be eclipsed by the NS (see Fig.\,\ref{pic:ColumnEcl}). The minimal height of a column which can be eclipsed is defined by mass and radius of NS. The eclipsing manifests itself by sharp periodic drops of X-ray flux within pulse profile. If eclipsing is detected in the pulse profile of a source we can get a lower limit on the NS radius.

\begin{figure}
\centering 
\includegraphics[width=8.5cm]{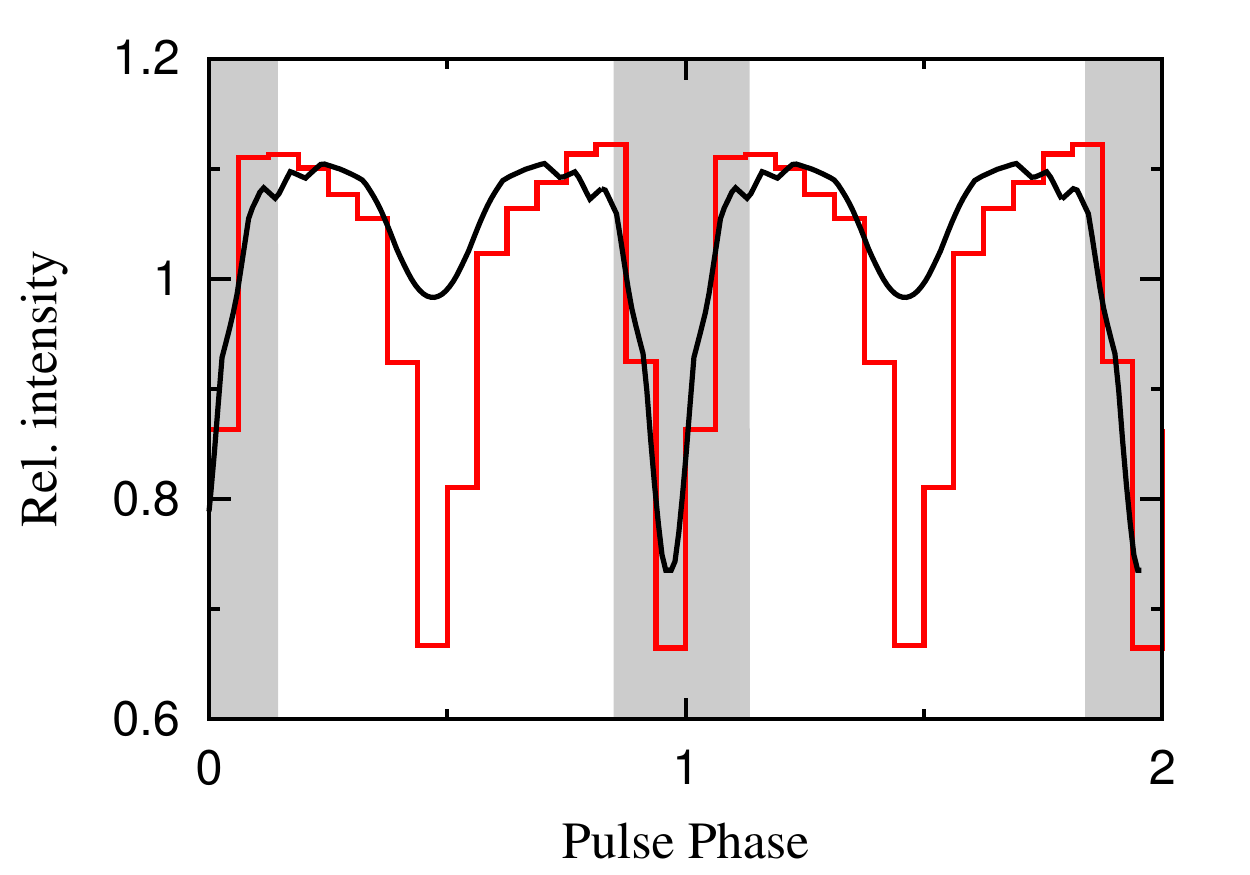} 	
\caption{An example of the pulse profile of V~0332+53 in its bright state of accretion luminosity $L=3.7\times 10^{37}\,\ergs$ (red line, see e.g. \citealt{2015MNRAS.448.2175L}) and theoretical pulse profile (black line). The X-ray flux at this luminosity is dominated by direct flux from accretion column, the sharp dips in the theoretical pulse profile (on top of grey colored phase range) are caused by eclipsing of the opposite accretion column by NS. The discrepancy between observed and theoretical pulse profiles at the phase $\sim 0.5$ is caused by oversimplified angular distribution of radiation at the accretion column walls. Parameters: $\mu=\eta=10^\circ$, $M=1.4M_\odot$, $R=15\,{\rm km}$, $H=9,3\,{\rm km}$.}
\label{pic:PP_comp}
\end{figure}

\begin{figure}
\centering 
\includegraphics[width=8.7cm]{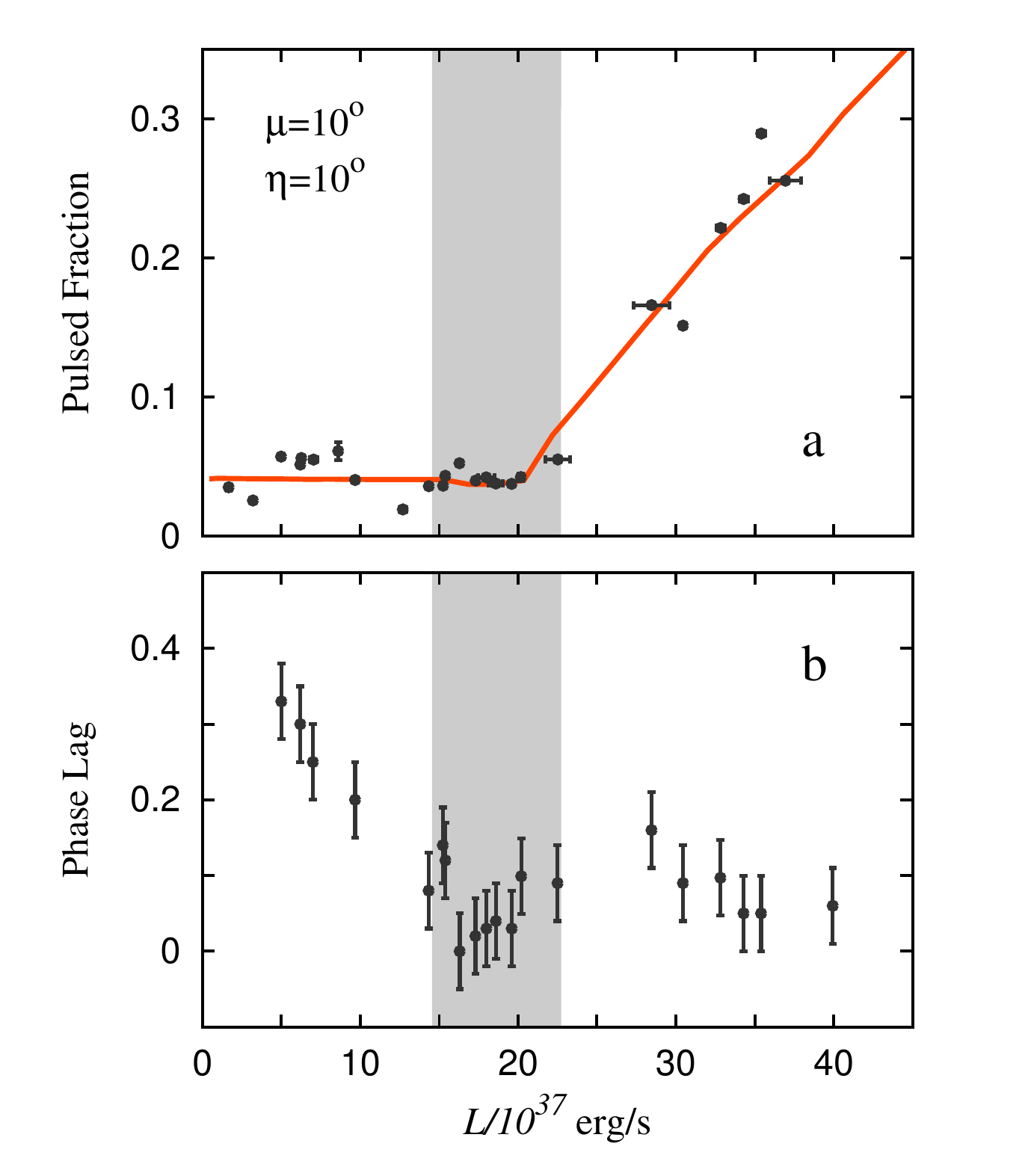} 	
\caption{(a) Dependence of the total pulsed fraction on the luminosity in the X-ray pulsar V~0332+53 obtained during the outburst 2004-2005 (circles with the error bars) and the theoretical dependence (solid red line) obtained for $\mu=\eta=10^\circ$, $M=1.4M_\odot$, $R=15\,{\rm km}$. 
(b) The phase lag between the maxim flux within the pulse profile and the phase of maximum centroid energy of the cyclotron scattering feature.
It is remarkable that the rapid increase of PF starts at about the luminosity where the phase lag between maximum flux in pulse profile and maximum cyclotron energy stabilizes at a relatively small value. 
}
\label{pic:PF}
\end{figure}

Such sharp drops of X-ray flux were detected in the super-critical XRP V~0332+53 in its high luminosity state (see Fig.\,\ref{pic:PP_comp} and also Fig.\,2 in \citealt{2015MNRAS.448.2175L}). No signs of strong scattering by matter in accretion channel were detected (we cannot exclude the possibility of scattering influence). 
The appearance of the drops of X-ray flux in V~0332+53 at high luminosity ($L\gtrsim 2\times 10^{38}\,\ergs$) strongly affects the pulsed fraction, which is almost constant at relatively low mass accretion rates but then rapidly increases with the luminosity from ${\rm PF}\sim 0.05$ at $L\sim 2\times 10^{38}\,\ergs$ up to ${\rm PF}\sim 0.35$ at $L\sim 4\times 10^{38}\,\ergs$ (see Fig.\,\ref{pic:PF}a).
\footnote{The PF dependence on accretion luminosity is taken from \cite{2010MNRAS.401.1628T}.}

This behavior of pulse profile and PF in V~0332+53 is naturally explained by eclipses of the accretion column: at low luminosity  the opposite column is not directly detectable, but at higher luminosity the column is high enough and becomes visible at some phases of the pulsation. Than the contribution of the column is larger, it provides the most X-ray flux at the maximum of  the pulse profile, while flux detected at the minimum of the pulse profile (during the phases of eclipsing, which are shown by grey regions on Fig.\,\ref{pic:PP_comp}) is dominated by reflected component.  
Because the eclipses are still detectable at accretion luminosity $L\sim 4\times 10^{38}\,\ergs$, when the accretion column height is expected to be comparable to the NS radius (see the approximate relation between column height and luminosity given by equation (\ref{eq:H2L})), we can get valuable lower limits on the NS radius in V~0332+53 (see Fig.\,\ref{pic:ColumnEcl}). Indeed, the observed pulsed fraction (and dips in pulse profiles due to eclipses, see model pulse profile on Fig.\,\ref{pic:PP_comp}) can be explained if the angles $\mu\approx\eta\approx 10^\circ$  (see Fig.\,\ref{pic:geom_scheme02}), the NS radius is taken to be $R\simeq 3.6 r_{\rm s}$ and the height of accretion column at maximum luminosity in the outburst is $H_{\rm max}\simeq 0.7 R$. For the case of NS mass $M=1.4M_\odot$ this corresponds to NS radius $R\sim 15\,{\rm km}$ and $H_{\rm max}\sim 10\,{\rm km}$.

It is interesting that the phase lag between the maximum in pulse profile and the phase where the centroid of the cyclotron scattering feature is at its maximum energy decreases with accretion luminosity and then stabilizes at a relatively small value at luminosity $L\sim 2\times 10^{38}\,\ergs$ (see Fig.\,\ref{pic:PF}b, \citealt{2015MNRAS.448.2175L}).
The cyclotron scattering feature forms at the NS surface due to reflection of the X-ray flux from the stellar atmosphere \citep{2013ApJ...777..115P} and varies with accretion luminosity and pulse phase. 
The maximum value of cyclotron centroid energy is detected when the NS magnetic pole is oriented towards the observer, because in this case the reprocessed flux is dominated by photons originating from regions close to the magnetic poles, where the surface $B$-field is stronger. Under these same conditions  the observer detects maximal X-ray flux, if the accretion columns are high enough \citep{2015MNRAS.448.2175L}. 
Thus, the observed stabilization of the phase lag at a small value is exactly what we expect at high luminosity. This supports our interpretation of the variations of the PF and pulse profiles in V~0332+53.

\begin{figure}
\centering 
\includegraphics[width=8.5cm]{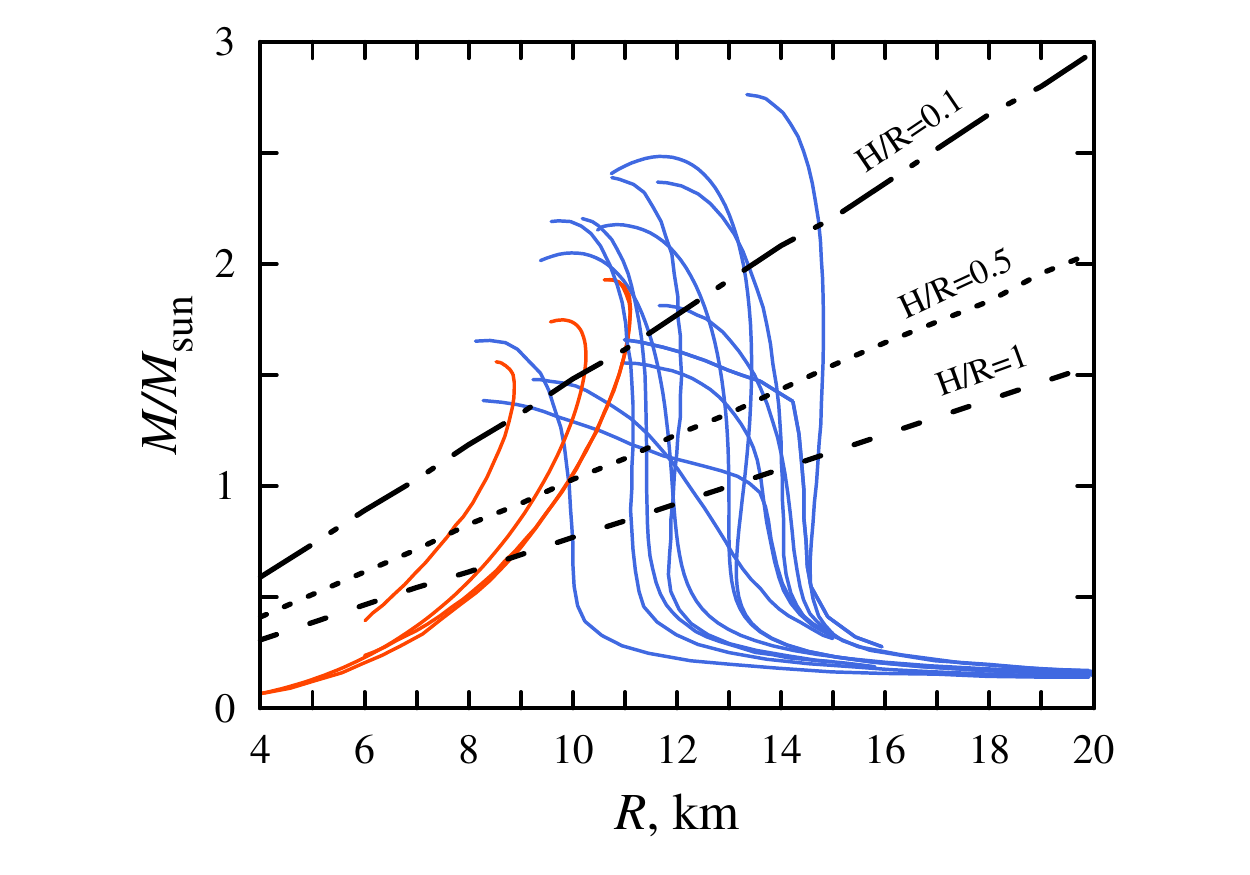} 	
\caption{Constrains on the NS mass-radius relation (blue lines correspond to NSs with various equations of states of dense matter, while red lines correspond to strange stars, see e.g. \citealt{2001ApJ...550..426L}), which can be obtained from eclipsing of accretion columns of different relative height $(H/R)$: the actual mass and radius should give a point located below corresponding limiting line.}
\label{pic:EoS}
\end{figure}

\subsection{Pulsating ULXs}

The recently discovered pulsating ULXs, whose accretion luminosity is known to be as high as $\sim 10^{40}-10^{41}\,\ergs$ \citep{2014Natur.514..202B,2017Sci...355..817I}, are the brightest accreting NSs known up to date. It very likely that their central engine is formed by accretion columns confined by a strong magnetic field \citep{2015MNRAS.454.2539M}. However, in case of ULXs powered by accreting NSs the accretion flow at the NS magnetosphere is optically thick and the central engine is hidden behind it. This explains the smooth pulse profiles of pulsating ULXs and predicts that the lensing/eclipsing features are not detectable in this class of accreting NSs.

\subsection{Sub-critical X-ray pulsars}

{At sub-critical mass accretion rates ($\dot{M}\lesssim 10^{17}\,{\rm g\,s^{-1}}$, see e.g \citealt{2015MNRAS.447.1847M}) the influence of radiation pressure is small and the accreting material is stopped either by Coulomb collisions in NS atmosphere \citep{1969SvA....13..175Z} or by a collisionless shock \citep{1975ApJ...198..671S}. Our knowledge of plasma physics in a strong magnetic field is insufficient to decide if a collisionless shock forms or not \citep{1982ApJ...257..733L,1987ApJ...312..666A}. However, if a collisionless shock forms (which is an assumption) then its height above the NS surface depends on the mass accretion rate: the lower the mass accretion rate, the higher the location of the collisionless shock. Numerical simulations show that the height of the shock can be $\sim 10^{5}\,{\rm cm}$ \citep{2004AstL...30..309B}.
Thus, if collisionless shocks form, one can expect the appearance of sharp dips due to eclipses of extended X-ray sources above the NS surface even in low luminosity ($L<10^{37}\,\ergs$) XRPs.
}

\section{Summary}
\label{discussion}

We have constructed a numerical model of radiation beaming from bright ($L\gtrsim 10^{37}\,\ergs$) XRPs, where the X-ray luminosity is generated in an accretion column. The total photon energy flux detected by a distant observer is composed of the direct X-ray flux from the accretion columns and the X-ray flux intercepted and reprocessed by the NS surface. The beam pattern, which defines the variability of the XRP over the pulse period, is strongly affected by effects of gravitational light bending, which is taken into account in this paper assuming Schwarzschild metric.

The detailed pulse profile is influenced by the exact initial beaming of X-ray flux at the edges of accretion column and the exact law of X-ray reflection from the NS atmosphere. However, there are a few qualitative aspects which are not dependent on the exact initial beaming and law of reflection. The direct flux from the column can be strongly amplified by gravitational lensing or reduced due to eclipsing by the NS in the direction opposite to accretion column. 
The exact scenario is determined by the accretion column height and the compactness of the central object. Eclipsing of the accretion column by the NS surface results in sharp dips in the pulse profile (see Fig.\,\ref{pic:PP_comp}). Appearance of dips in the pulse profiles is accompanied by a fast increase of the PF with luminosity and accretion column height. The X-ray flux is expected to be dominated by the direct flux from the accretion column in the luminosity range where the PF rapidly increases. 
The NS can eclipse the accretion column only in the case of a sufficiently large radius. Thus, observational detection of eclipsing in an XRP based on the pulse profile and the behavior of the PF at given mass accretion rate (and therefore given height of accretion column) gives an opportunity to put a constraint on the NS radius (see Fig.\,\ref{pic:EoS} and Fig.\,\ref{pic:ColumnEcl}): 
\be
R\gtrsim 6\,\left(1+\frac{7}{6}\frac{H}{R}\right) \left(\frac{M}{M_{\odot}}\right)\,\,{\rm km},
\ee
where $H/R\in [0.1,1]$ is relative height of accretion column eclipsed by NS. The relative height at given mass accretion rate should be provided by the model of the accretion column.	

We have applied our model to the bright XRP V~0332+53 and reconstructed the observed dependence of PF on accretion luminosity (see Fig.\,\ref{pic:PF}a). According to our results the rotational and magnetic field axis in this particular source are slightly inclined to the line of sight ($\mu\sim 10^\circ$, see Fig.\,\ref{pic:geom_scheme02}). If this is a case, the NS in V~0332+53 is a good candidate to detect the accretion column eclipsing. We speculate, that the sharp dips detected in the pulse profiles of V~0332+53 at luminosity $L\sim 3.7\times 10^{38}\,\ergs$ \citep{2015MNRAS.448.2175L} can be caused by eclipsing. The behavior of the cyclotron line scattering feature in the phase-resolved spectra (see Fig.\,\ref{pic:PF}b) supports our interpretation.
If the sharp dips in the pulse profile are caused by eclipsing of the accretion column, then the NS radius in this particular system has to be large: $\sim 15\,{\rm km}$. 

We point out, that the influence of the accretion flow at the NS magnetosphere can be crucial in the formation of pulse profiles of bright XRPs. It is likely determined by Compton scattering, which is resonant for X-ray photons in the vicinity of a magnetized NS. The resonant scattering occurs at different heights depending on photon energy and magnetic field strength and structure. It disturbs the X-ray pulse profiles at energies below the cyclotron energy at the NS surface. At the extreme accretion luminosities typical for pulsating ULXs ($\gtrsim 10^{40}\,\ergs$) the accretion envelope at the magnetosphere is optically thick, which shapes the pulse profiles of accreting NSs \citep{2017MNRAS.467.1202M}. In case of high accretion columns ($H\sim R$) above compact NS ($R\lesssim 2.5r_{\rm s}$), the accretion flow can be influenced by the X-ray flux from the opposite accretion column.

\section*{Acknowledgements}

AAM and MvdK thanks for support by the Netherlands Organization for Scientific Research (NWO) in part of construction of the basic theoretical model.
AAM, SST and AAL also acknowledge support by the Russian Science Foundation grant 14-12-01287 in part of critical analysis of the observational data and fitting the data by proposed model (Sections 5.1).
We are grateful to Dmitry Yakovlev, Yuri Shibanov and Alexander Potekhin for a number of useful comments.


\begin{thebibliography}{}

\bibitem[\protect\citeauthoryear{{Annala} \& {Poutanen}}{{Annala} \&
  {Poutanen}}{2010}]{2010A&A...520A..76A}
{Annala} M.,  {Poutanen} J.,  2010, \aap, 520, A76

\bibitem[\protect\citeauthoryear{{Arons}, {Klein} \& {Lea}}{{Arons}
  et~al.}{1987}]{1987ApJ...312..666A}
{Arons} J.,  {Klein} R.~I.,    {Lea} S.~M.,  1987, \apj, 312, 666

\bibitem[\protect\citeauthoryear{{Bachetti}}{{Bachetti}}{2014}]{2014Natur.514..202B}
{Bachetti} M. et al.,  2014, \nat, 514, 202

\bibitem[\protect\citeauthoryear{{Basko} \& {Sunyaev}}{{Basko} \&
  {Sunyaev}}{1975}]{1975A&A....42..311B}
{Basko} M.~M.,  {Sunyaev} R.~A.,  1975, \aap, 42, 311

\bibitem[\protect\citeauthoryear{{Basko} \& {Sunyaev}}{{Basko} \&
  {Sunyaev}}{1976}]{BS1976}
{Basko} M.~M.,  {Sunyaev} R.~A.,  1976, \mnras, 175, 395

\bibitem[\protect\citeauthoryear{{Beloborodov}}{{Beloborodov}}{2002}]{2002ApJ...566L..85B}
{Beloborodov} A.~M.,  2002, \apjl, 566, L85

\bibitem[\protect\citeauthoryear{{Bykov} \& {Krasil'shchikov}}{{Bykov} \&
  {Krasil'shchikov}}{2004}]{2004AstL...30..309B}
{Bykov} A.~M.,  {Krasil'shchikov} A.~M.,  2004, Astronomy Letters, 30, 309

\bibitem[\protect\citeauthoryear{{Doroshenko}, {Santangelo}, {Doroshenko},
  {Suleimanov} \& {Piraino}}{{Doroshenko} et~al.}{2015}]{2015MNRAS.452.2490D}
{Doroshenko} R.,  {Santangelo} A.,  {Doroshenko} V.,  {Suleimanov} V.,
  {Piraino} S.,  2015, \mnras, 452, 2490

\bibitem[\protect\citeauthoryear{{Doroshenko}, {Tsygankov}, {Mushtukov},
  {Lutovinov}, {Santangelo}, {Suleimanov} \& {Poutanen}}{{Doroshenko}
  et~al.}{2017}]{2017MNRAS.466.2143D}
{Doroshenko} V. et al.,  2017, \mnras, 466,
  2143

\bibitem[\protect\citeauthoryear{{Gnedin} \& {Sunyaev}}{{Gnedin} \&
  {Sunyaev}}{1973}]{1973A&A....25..233G}
{Gnedin} Y.~N.,  {Sunyaev} R.~A.,  1973, \aap, 25, 233

\bibitem[\protect\citeauthoryear{{Harding} \& {Lai}}{{Harding} \&
  {Lai}}{2006}]{2006RPPh...69.2631H}
{Harding} A.~K.,  {Lai} D.,  2006, Reports on Progress in Physics, 69, 2631

\bibitem[\protect\citeauthoryear{{Herold}}{{Herold}}{1979}]{1979PhRvD..19.2868H}
{Herold} H.,  1979, \prd, 19, 2868

\bibitem[\protect\citeauthoryear{{Israel}}{{Israel}}{2017a}]{2017Sci...355..817I}
{Israel} G.~L. et al.,  2017a, Science, 355, 817

\bibitem[\protect\citeauthoryear{{Israel}}{{Israel}}{2017b}]{2017MNRAS.466L..48I}
{Israel} G.~L. et al.,  2017b, \mnras, 466, L48

\bibitem[\protect\citeauthoryear{{Kaminker}, {Fedorenko} \&
  {Tsygan}}{{Kaminker} et~al.}{1976}]{1976SvA....20..436K}
{Kaminker} A.~D.,  {Fedorenko} V.~N.,    {Tsygan} A.~I.,  1976, \sovast, 20,
  436

\bibitem[\protect\citeauthoryear{{Kraus}}{{Kraus}}{2001}]{2001ApJ...563..289K}
{Kraus} U.,  2001, \apj, 563, 289

\bibitem[\protect\citeauthoryear{{Kraus}, {Nollert}, {Ruder} \&
  {Riffert}}{{Kraus} et~al.}{1995}]{1995ApJ...450..763K}
{Kraus} U.,  {Nollert} H.-P.,  {Ruder} H.,    {Riffert} H.,  1995, \apj, 450,
  763

\bibitem[\protect\citeauthoryear{{Lai}}{{Lai}}{2014}]{2014EPJWC..6401001L}
{Lai} D.,  2014, in European Physical Journal Web of Conferences Vol.~64 of
  European Physical Journal Web of Conferences, {Theory of Disk Accretion onto
  Magnetic Stars}.
p. 01001

\bibitem[\protect\citeauthoryear{{Langer} \& {Rappaport}}{{Langer} \&
  {Rappaport}}{1982}]{1982ApJ...257..733L}
{Langer} S.~H.,  {Rappaport} S.,  1982, \apj, 257, 733

\bibitem[\protect\citeauthoryear{{Lattimer} \& {Prakash}}{{Lattimer} \&
  {Prakash}}{2001}]{2001ApJ...550..426L}
{Lattimer} J.~M.,  {Prakash} M.,  2001, \apj, 550, 426

\bibitem[\protect\citeauthoryear{{Lutovinov}, {Tsygankov}, {Suleimanov},
  {Mushtukov}, {Doroshenko}, {Nagirner} \& {Poutanen}}{{Lutovinov}
  et~al.}{2015}]{2015MNRAS.448.2175L}
{Lutovinov} A.~A. et al.,  2015, \mnras, 448,
  2175

\bibitem[\protect\citeauthoryear{{Lyubarskii} \& {Syunyaev}}{{Lyubarskii} \&
  {Syunyaev}}{1988}]{1988SvAL...14..390L}
{Lyubarskii} Y.~E.,  {Syunyaev} R.~A.,  1988, Soviet Astronomy Letters, 14, 390

\bibitem[\protect\citeauthoryear{{Misner}, {Thorne} \& {Wheeler}}{{Misner}
  et~al.}{1973}]{1973grav.book.....M}
{Misner} C.~W.,  {Thorne} K.~S.,    {Wheeler} J.~A.,  1973, {Gravitation}

\bibitem[\protect\citeauthoryear{{Mitrofanov} \& {Tsygan}}{{Mitrofanov} \&
  {Tsygan}}{1978}]{1978A&A....70..133M}
{Mitrofanov} I.~G.,  {Tsygan} A.~I.,  1978, \aap, 70, 133

\bibitem[\protect\citeauthoryear{{Mukherjee}, {Bhattacharya} \&
  {Mignone}}{{Mukherjee} et~al.}{2013a}]{2013MNRAS.430.1976M}
{Mukherjee} D.,  {Bhattacharya} D.,    {Mignone} A.,  2013a, \mnras, 430, 1976

\bibitem[\protect\citeauthoryear{{Mukherjee}, {Bhattacharya} \&
  {Mignone}}{{Mukherjee} et~al.}{2013b}]{2013MNRAS.435..718M}
{Mukherjee} D.,  {Bhattacharya} D.,    {Mignone} A.,  2013b, \mnras, 435, 718

\bibitem[\protect\citeauthoryear{{Mushtukov}, {Nagirner} \&
  {Poutanen}}{{Mushtukov} et~al.}{2016}]{2016PhRvD..93j5003M}
{Mushtukov} A.~A.,  {Nagirner} D.~I.,    {Poutanen} J.,  2016, \prd, 93, 105003

\bibitem[\protect\citeauthoryear{{Mushtukov}, {Suleimanov}, {Tsygankov} \&
  {Ingram}}{{Mushtukov} et~al.}{2017}]{2017MNRAS.467.1202M}
{Mushtukov} A.~A.,  {Suleimanov} V.~F.,  {Tsygankov} S.~S.,    {Ingram} A.,
  2017, \mnras, 467, 1202

\bibitem[\protect\citeauthoryear{{Mushtukov}, {Suleimanov}, {Tsygankov} \&
  {Poutanen}}{{Mushtukov} et~al.}{2015a}]{2015MNRAS.454.2539M}
{Mushtukov} A.~A.,  {Suleimanov} V.~F.,  {Tsygankov} S.~S.,    {Poutanen} J.,
  2015a, \mnras, 454, 2539

\bibitem[\protect\citeauthoryear{{Mushtukov}, {Suleimanov}, {Tsygankov} \&
  {Poutanen}}{{Mushtukov} et~al.}{2015b}]{2015MNRAS.447.1847M}
{Mushtukov} A.~A.,  {Suleimanov} V.~F.,  {Tsygankov} S.~S.,    {Poutanen} J.,
  2015b, \mnras, 447, 1847

\bibitem[\protect\citeauthoryear{{Mushtukov}, {Tsygankov}, {Serber},
  {Suleimanov} \& {Poutanen}}{{Mushtukov} et~al.}{2015}]{2015MNRAS.454.2714M}
{Mushtukov} A.~A.,  {Tsygankov} S.~S.,  {Serber} A.~V.,  {Suleimanov} V.~F.,
  {Poutanen} J.,  2015, \mnras, 454, 2714

\bibitem[\protect\citeauthoryear{{Pechenick}, {Ftaclas} \& {Cohen}}{{Pechenick}
  et~al.}{1983}]{1983ApJ...274..846P}
{Pechenick} K.~R.,  {Ftaclas} C.,    {Cohen} J.~M.,  1983, \apj, 274, 846

\bibitem[\protect\citeauthoryear{{Postnov}, {Gornostaev}, {Klochkov},
  {Laplace}, {Lukin} \& {Shakura}}{{Postnov}
  et~al.}{2015}]{2015MNRAS.452.1601P}
{Postnov} K.~A.,  {Gornostaev} M.~I.,  {Klochkov} D.,  {Laplace} E.,  {Lukin}
  V.~V.,    {Shakura} N.~I.,  2015, \mnras, 452, 1601

\bibitem[\protect\citeauthoryear{{Potekhin}}{{Potekhin}}{2014}]{2014PhyU...57..735P}
{Potekhin} A.~Y.,  2014, Physics Uspekhi, 57, 735

\bibitem[\protect\citeauthoryear{{Poutanen} \& {Beloborodov}}{{Poutanen} \&
  {Beloborodov}}{2006}]{2006MNRAS.373..836P}
{Poutanen} J.,  {Beloborodov} A.~M.,  2006, \mnras, 373, 836

\bibitem[\protect\citeauthoryear{{Poutanen}, {Mushtukov}, {Suleimanov},
  {Tsygankov}, {Nagirner}, {Doroshenko} \& {Lutovinov}}{{Poutanen}
  et~al.}{2013}]{2013ApJ...777..115P}
{Poutanen} J. et al.,  2013, \apj, 777,
  115

\bibitem[\protect\citeauthoryear{{Riffert} \& {Meszaros}}{{Riffert} \&
  {Meszaros}}{1988}]{1988ApJ...325..207R}
{Riffert} H.,  {Meszaros} P.,  1988, \apj, 325, 207

\bibitem[\protect\citeauthoryear{{Sasaki}, {M{\"u}ller}, {Kraus}, {Ferrigno} \&
  {Santangelo}}{{Sasaki} et~al.}{2012}]{2012A&A...540A..35S}
{Sasaki} M.,  {M{\"u}ller} D.,  {Kraus} U.,  {Ferrigno} C.,    {Santangelo} A.,
   2012, \aap, 540, A35

\bibitem[\protect\citeauthoryear{{Shapiro} \& {Salpeter}}{{Shapiro} \&
  {Salpeter}}{1975}]{1975ApJ...198..671S}
{Shapiro} S.~L.,  {Salpeter} E.~E.,  1975, \apj, 198, 671

\bibitem[\protect\citeauthoryear{{Syunyaev}}{{Syunyaev}}{1976}]{1976SvAL....2..111S}
{Syunyaev} R.~A.,  1976, Soviet Astronomy Letters, 2, 111

\bibitem[\protect\citeauthoryear{{Tsygankov}, {Doroshenko}, {Lutovinov},
  {Mushtukov} \& {Poutanen}}{{Tsygankov} et~al.}{2017}]{2017A&A...605A..39T}
{Tsygankov} S.~S.,  {Doroshenko} V.,  {Lutovinov} A.~A.,  {Mushtukov} A.~A.,
  {Poutanen} J.,  2017, \aap, 605, A39

\bibitem[\protect\citeauthoryear{{Tsygankov}, {Krivonos} \&
  {Lutovinov}}{{Tsygankov} et~al.}{2012}]{2012MNRAS.421.2407T}
{Tsygankov} S.~S.,  {Krivonos} R.~A.,    {Lutovinov} A.~A.,  2012, \mnras, 421,
  2407

\bibitem[\protect\citeauthoryear{{Tsygankov}, {Lutovinov}, {Churazov} \&
  {Sunyaev}}{{Tsygankov} et~al.}{2006}]{2006MNRAS.371...19T}
{Tsygankov} S.~S.,  {Lutovinov} A.~A.,  {Churazov} E.~M.,    {Sunyaev} R.~A.,
  2006, \mnras, 371, 19

\bibitem[\protect\citeauthoryear{{Tsygankov}, {Lutovinov}, {Doroshenko},
  {Mushtukov}, {Suleimanov} \& {Poutanen}}{{Tsygankov}
  et~al.}{2016}]{2016A&A...593A..16T}
{Tsygankov} S.~S.,  {Lutovinov} A.~A.,  {Doroshenko} V.,  {Mushtukov} A.~A.,
  {Suleimanov} V.,    {Poutanen} J.,  2016, \aap, 593, A16

\bibitem[\protect\citeauthoryear{{Tsygankov}, {Lutovinov} \&
  {Serber}}{{Tsygankov} et~al.}{2010}]{2010MNRAS.401.1628T}
{Tsygankov} S.~S.,  {Lutovinov} A.~A.,    {Serber} A.~V.,  2010, \mnras, 401,
  1628

\bibitem[\protect\citeauthoryear{{Walter}, {Lutovinov}, {Bozzo} \&
  {Tsygankov}}{{Walter} et~al.}{2015}]{2015A&ARv..23....2W}
{Walter} R.,  {Lutovinov} A.~A.,  {Bozzo} E.,    {Tsygankov} S.~S.,  2015,
  \aapr, 23, 2

\bibitem[\protect\citeauthoryear{{Zel'dovich} \& {Shakura}}{{Zel'dovich} \&
  {Shakura}}{1969}]{1969SvA....13..175Z}
{Zel'dovich} Y.~B.,  {Shakura} N.~I.,  1969, \sovast, 13, 175

\bibitem[\protect\citeauthoryear{{Zheleznyakov} \& {Litvinchuk}}{{Zheleznyakov}
  \& {Litvinchuk}}{1986}]{1986ESASP.251..375Z}
{Zheleznyakov} V.~V.,  {Litvinchuk} A.~A.,  1986, in {Guyenne} T.~D.,  {Zeleny}
  L.~M.,  eds, Plasma Astrophysics Vol.~251 of ESA Special Publication,
  {Radiation transfer and radiation pressure on plasma by magnetic degenerates}

\end{thebibliography}

{

}

\bsp	
\label{lastpage}
\end{document}